\documentclass[twocolumn,twocolappendix]{aastex63}
\usepackage{subfigure}
\usepackage{array}
\def\iso#1{$^{#1}$}

\received{2022 April 28}
\revised{2022 December 31}
\accepted{2023 January 02}
\published{2023 February 16}
\shorttitle{SLRs: Surfing the wave?}
\shortauthors{Wehmeyer et al.}
\begin{document}
\title{Inhomogeneous enrichment of radioactive nuclei in the Galaxy: Deposition of live \iso{53}Mn, \iso{60}Fe, \iso{182}Hf, and \iso{244}Pu into deep-sea archives. Surfing the wave?}
\author{Benjamin Wehmeyer}
\affiliation{Konkoly Observatory, Research Centre for Astronomy and Earth Sciences, E\"otv\"os Lor\'and Research Network (ELKH),\\ Excellence Centre of the Hungarian Academy of Sciences,\\ Konkoly-Thege Mikl\'os \'ut 15-17, H-1121 Budapest, Hungary}
\affiliation{Centre for Astrophysics Research, University of Hertfordshire,\\ College Lane, Hatfield AL10 9AB, UK}
\author[0000-0002-7294-9288]{Andr\'es Yag\"ue L\'opez}
\affiliation{Konkoly Observatory, Research Centre for Astronomy and Earth Sciences, E\"otv\"os Lor\'and Research Network (ELKH),\\ Excellence Centre of the Hungarian Academy of Sciences,\\ Konkoly-Thege Mikl\'os \'ut 15-17, H-1121 Budapest, Hungary}
\affiliation{Computer, Computational and Statistical Sciences (CCS) Division, Center for Theoretical Astrophysics, Los Alamos National Laboratory, Los Alamos, NM 87545, USA}
\author[0000-0002-9986-8816]{Benoit C\^ot\'e}
\affiliation{Department of Physics and Astronomy, University of Victoria, BC, V8W 2Y2, Canada}
\affiliation{Konkoly Observatory, Research Centre for Astronomy and Earth Sciences, E\"otv\"os Lor\'and Research Network (ELKH),\\ Excellence Centre of the Hungarian Academy of Sciences,\\ Konkoly-Thege Mikl\'os \'ut 15-17, H-1121 Budapest, Hungary}
\author{Maria K. Pet\H{o}}
\affiliation{Konkoly Observatory, Research Centre for Astronomy and Earth Sciences, E\"otv\"os Lor\'and Research Network (ELKH),\\ Excellence Centre of the Hungarian Academy of Sciences,\\ Konkoly-Thege Mikl\'os \'ut 15-17, H-1121 Budapest, Hungary}
\author[0000-0002-4343-0487]{Chiaki Kobayashi}
\affiliation{Centre for Astrophysics Research, University of Hertfordshire,\\ College Lane, Hatfield AL10 9AB, UK}
\author[0000-0002-6972-3958]{Maria Lugaro}
\affiliation{Konkoly Observatory, Research Centre for Astronomy and Earth Sciences, E\"otv\"os Lor\'and Research Network (ELKH),\\ Excellence Centre of the Hungarian Academy of Sciences,\\ Konkoly-Thege Mikl\'os \'ut 15-17, H-1121 Budapest, Hungary}
\affiliation{ELTE E\"otv\"os Lor\'and University, Institute of Physics, Budapest 1117, Hungary}
\affiliation{School of Physics and Astronomy, Monash University, VIC 3800, Australia}
\correspondingauthor{Benjamin Wehmeyer}
\email{benjamin.wehmeyer@csfk.org}
\begin{abstract}
While modelling the galactic chemical evolution (GCE) of stable elements provides insights to the formation history of the Galaxy and the relative contributions of nucleosynthesis sites, modelling the evolution of short-lived radioisotopes (SLRs) can provide supplementary timing information on recent nucleosynthesis. To study the evolution of SLRs, we need to understand their spatial distribution. Using a 3-dimensional GCE model, we investigated the evolution of four SLRs: \iso{53}Mn, \iso{60}Fe, \iso{182}Hf, and \iso{244}Pu with the aim of explaining detections of recent (within the last $\approx 1-20$ Myr) deposition of live \iso{53}Mn, \iso{60}Fe, and \iso{244}Pu of extrasolar origin into deep-sea reservoirs. 
We find that core-collapse supernovae (CCSNe) are the dominant propagation mechanism of SLRs in the Galaxy. This results in the simultaneously arrival of these four SLRs on Earth, although they could have been produced in different astrophysical sites, which can  explain why live extrasolar \iso{53}Mn, \iso{60}Fe, and \iso{244}Pu are found within the same, or similar, layers of deep-sea sediments. We predict that \iso{182}Hf should also be found in such sediments at similar depths.
\end{abstract}
\section{Introduction} \label{sec:intro}
Studying the galactic chemical evolution (GCE) of stable elements provides insights on the formation of the Galaxy and the nucleosynthesis processes that produced the chemical elements \citep[e.g.,][]{Audouze76,Matteucci86,gibson03,Nomoto13,Kobayashi20,prantzos20}. The abundance of short-lived radioactive isotopes (SLRs), with half-lives of less than a few 100 Myrs, instead follows the trend of star formation rate, and is determined by the balance between their production and decay \citep[the steady-state equilibrium, e.g.,][]{Clayton84}.
SLRs can be observed live in the interstellar medium \citep[ISM, e.g.,][]{Diehl10}, in Earth deep-sea sediments accumulating in the last 10 Myr with a constant rate \citep{Wallner15,Wallner16,Korschinek20,Wallner21} and extinct in meteorites \citep{dauphas11,Lugaro18a}, and offer additional information on the timing of processes and our Solar System. Comparing the SLR abundances in meteorites to the abundances predicted by GCE models can constrain the last stellar events contributing SLRs to the molecular cloud in which the Solar System formed, and determine the time the Solar System matter remained isolated from the Galactic medium into such molecular cloud \citep[e.g.,][]{Clayton84,Meyer00,Huss09,Lugaro14,Lugaro18a,Cote19a,cotescience,Trueman22}.
Furthermore, comparing SLR abundances in the ISM at any two points in time (e.g., at the time of Solar System formation from meteorites and today from deep-sea sediments) to GCE models can allow us to constrain the origin of a given SLR based on differences in the event rates of stellar processes producing such a given SLR \citep[e.g.,][]{Hotokezaka15}.
In addition, the SLR abundances are reported in deep-sea sediments with high resolution temporal profiles (with 200 kyr sampling), which showed live \iso{60}Fe (half-life t$_{1/2}=2.62$ Myr) originating from the ISM together with resolvable \iso{244}Pu (t$_{1/2}=80.0$ Myr) and \iso{53}Mn (t$_{1/2}=3.74$ Myr) anomalies  \citep[e.g.,][]{Wallner15,Wallner16,Korschinek20,Wallner21}. These profiles identified at least 2 distinct signals of SLR deposition to Earth, $2.5 \pm 0.5$ and at $5.4 \pm 0.7$ Myr ago. The temporal overlap of the \iso{53}Mn and \iso{60}Fe signals supports the involvement of a core-collapse supernova (CCSN) origin of these signals, while the temporal overlap with \iso{244}Pu signals may indicate a rare supernova event or a more complicated transport history of material in the ISM \citep[e.g.,][]{Hotokezaka15}.  
To best interpret all these observations, we need to better understand the dependence of SLR abundances on GCE model parameters, the temporal and spatial stochasticity of enrichment events from the different stellar sources, and the transport mechanism of material in the ISM.
To accommodate SLRs in more recent GCE models, \citet[][]{Cote19a} derived the behaviour of the ratio of SLRs to their stable reference isotopes over the lifetime of the Galaxy using the two-zone GCE model \textsc{Omega+} \citep{omegaplus}. This study explained and quantified  the effects of galactic inflows and outflows, delay-time distributions of enrichment events from different stellar sources, the Galactic star formation history, and the gas-to-star mass ratio, assuming a continuous production rate of SLRs and a homogeneous ISM. 
This quantification of uncertainties on the average composition of GCE models was recently extended to consider inhomogeneities in the ISM due to the temporal stochasticity of stellar enrichment events. \citet[][]{Cote19b} developed a Monte Carlo simulation for the temporal evolution of SLRs in a local region of the interstellar gas.  The study developed a general statistical framework to quantify the uncertainty (probability distributions) of SLR abundances based on the stochastic delay time between star formation and enrichment, and explained the dependence of SLR abundances on the average time between enriching events, the delay time distribution, and the mean life of the SLRs. \cite{Cote19b} also quantified the probability of whether an SLR could sample only one stellar event or whether a particular SLR abundance represents a steady-state in the ISM, where frequent production of nuclides is balanced out by their decay.
\citet[][]{Yague21} studied the abundance ratios of two SLRs using the same statistical framework, which also depend on the relative mean life of the SLRs. 
The ratios of SLRs with similar, and short enough half-lives are largely independent of GCE model uncertainties, and therefore are unique tools to study the nucleosynthesis at their formation sites, provided that their production occurs synchronously. The theoretical considerations in \citet[][]{Yague21} have been successfully applied to the \iso{129}I/\iso{247}Cm ratio to understand the physical condition of the last rapid neutron capture process event that contributed to the Solar System inventory prior to its formation \citep[][]{cotescience}.
The focus of the present study is to use a full inhomogeneous GCE code to consider not only temporal discretization (as done in the papers mentioned above) but also spatial discretization of SLRs and the role of inhomogeneities in the ISM to develop a better understanding of transport of matter in our Galaxy. In particular, we aim to explain the possibly synchronous delivery of live \iso{53}Mn, \iso{60}Fe, and \iso{244}Pu onto the ocean floor as recorded in the deep-sea sediments deposited in the last 10 Myr. So far, it has been difficult to draw conclusions about the impact of these detections using GCE models that assume homogeneous mixing, since spatial discretization effects have a significant impact on the abundances of the detected SLRs.
Since these problems are difficult or impossible to study with 1-dimensional models, here we use the 3-dimensional GCE model from \cite{Wehmeyer19}. We simulate the spatial and temporal evolution of four SLR abundances over the lifetime of the Galaxy: \iso{53}Mn, \iso{60}Fe, \iso{182}Hf (t$_{1/2}=8.90$ Myr), and \iso{244}Pu to draw conclusions about their most recent and ongoing (within the last Myrs) deposition to Earth. The comparison of our model results to SLR abundances in the early Solar System derived from meteorites and its implications will be published in a separate study. 
This paper is organized as follows. In Section~\ref{sec:model}, we summarize the main modeling parameters. In Section~\ref{sec:Results:Evolution of SLRs} we discuss the abundance evolution of the selected SLRs over the entire lifetime of the Galaxy. In Section~\ref{sec:results:recentevolution}, we zoom in on the more recent time period closer to present day, and compare our simulation results with deep-sea sediment detections. In Sections~\ref{sec:results:Pollution intervals} and \ref{sec:results:dominant prop mechanism}, we compare the propagation of the different SLRs in the simulation volume and derive the dominant transport mechanism for SLRs. We present a schematic interpretation of our findings in Section~\ref{sec:schematic_interpretation}, discuss the impact of yields in Section~\ref{sec:results:evolutionof SLRS:Yields} and provide our conclusions in Section~\ref{sec:conclusions}.
\section{The model} \label{sec:model}
For this study, we use the 3-dimensional GCE model described in \cite{Wehmeyer15,Wehmeyer19}. Below, we recall the most important modelling assumptions, and highlight relevant updates to the model.
\subsection{General setup}\label{Sec:Model:General setup}
A simulation cube of $(2 \mathrm{kpc)^3}$ is divided into $40^3$ sub-cubes (or cells) with an edge length of 50 pc each.
During each calculated time step of 1\,Myr, the following operations are performed.
\begin{enumerate}
    \item Gas with primordial composition falls into the simulation volume according to the prescription of \cite{Wehmeyer15}, which permits for a linear rise of infalling material until 2\,Gyr, and then an exponential decrease of the infall rate. When falling into the simulation volume, the gas is homogeneously distributed among all sub-cubes.
    \item The total gas mass of the simulation volume is used to determine the number of stars to be born based on a Schmidt law with power $\alpha=1.5$ (\citealt{Schmidt59,Kennicutt98,Larson91}). The mass of the new born stars is sampled from a Salpeter initial mass function (IMF) \citep[][]{Salpeter55} with a slope of $-2.35$, and mass limits of 0.1 $\mathrm{M}_\odot \leq \mathrm{M} < $ 1 $\mathrm{M}_\odot$\footnote{Concerning star masses, we refer to the zero age main sequence (ZAMS) mass of the star throughout this manuscript.} for low-mass stars (LMS), 1 $\mathrm{M}_\odot \leq \mathrm{M} < $ 10 $\mathrm{M}_\odot$ for intermediate-mass stars (IMS), and 10 $\mathrm{M}_\odot \leq \mathrm{M} \leq 50 \mathrm{M}_\odot$ for high-mass stars (HMS). The newly born stars inherit the chemical composition of the gas out of which they were formed.
    \item Once the number and masses of the new born stars are known, their birth location is chosen randomly. Due to supernova explosions (Section~\ref{Sec:Model:HMS}), the gas density distribution becomes more and more inhomogeneous with time (i.e., the density distribution function steepens), and preference of star formation is given to cells with higher gas densities due to the Schmidt law. To prevent missampling of the IMF, only cells containing at least  50 $\mathrm{M}_\odot$ of gas are permitted to form stars. This constraint is also a limiting factor to increase the resolution of the sub-cells: every time step, a sufficiently large number of sub-cells has to be available for star formation, i.e., has to fulfil this minimum mass requirement. If the resolution of the sub-cells was increased, not enough sub-cells available for star formation would be found during a time step. This would alter the star formation rate and lead to other problems (e.g., with the applicability of the model on the GCE of $\alpha$-elements). Decreasing the spatial resolution (i.e., using larger sub-cell size) is instead feasible and we tested the dependence of our model on this in Appendix~\ref{appendix:sec:resolution}: We find that decreasing the sub-cell resolution results in the abundance spectrum getting smaller (i.e., converging towards a line), which resembles rather a one-zone model behavior.
    \item The lifetime for every newly born star is calculated using the formula by the Geneva group (see \citealt{Schaller92,Schaerer93b,Schaerer93a,Charbonnel93}):
    \begin{equation}\label{Model:Lifetimeformula}
\begin{split}
    \text{log}(t)&=(3.79 + 0.24  Z) - (3.10 + 0.35 Z)  \text{log} (M) \\
    & + (0.74 + 0.11 Z)  \text{log}^2(M)\text{,}
            \end{split}
    \end{equation}
    where $t$ is the expected lifetime of a star in Myr, $Z$ is the metallicity with respect to solar, and $\mathrm{M}$ the stellar mass in solar masses.
    \item If a time step contains stars that have reached the end of their lifetime, stellar death is simulated following the description detailed in the next section.
\end{enumerate}
\subsection{Stellar deaths} \label{Sec:Model:Nucleosynthesis sites}
\subsubsection{LMS and IMS} \label{Sec:Model:LIMS}
During their lifetime, LMS \& IMS add significant amounts of C and N to the galactic inventory \citep[e.g.,][]{Kobayashi11agb}. They do not reach burning stages more advanced than He burning, and thus do not contribute to iron-group elements significantly, however they produce slow neutron capture (s-process) elements such as Sr, Y, Zr, Ba, La, Ce, and Pb, as well as Hf of interest here during their thermally-pulsing asymptotic giant branch (AGB) phase \citep[e.g.,][]{Kappeler11,Bisterzo14,Kobayashi20}.
When LMS \& IMS die, they eject some of their initial abundances, plus their nucleosynthesis products via stellar winds, except for the fraction that remains locked in the degenerate core. Their death may result in a planetary nebula, and then a white dwarf remnant, and it is far less violent/energetic compared to the death of a HMS because it is wind-driven instead of explosion-driven. The ejecta of IMS typically pollute only volumes with diameters of the order of light years (e.g., the Cat's eye nebula NGC 6543 has a radius of 0.1 light year, \citealt{Reed99}, and the Helix nebula NGC 7293 a radius of 1.43 light years, \citealt{Odell04}). LMS \& IMS have two main functions in our simulation: to lock up gas during their lifetime, and to produce heavy isotopes via the s-process, including \iso{182}Hf.
These isotopes are injected at the location of the source, since the energy injection by the site is negligible.
\subsubsection{Thermonuclear supernovae}
Since many stars are born in binary systems (e.g., \citealt{Duchene13}), a significant fraction of IMS interact with a companion and undergo a supernova of type Ia (SNIa), which are the dominant source of Fe in the galactic disk \citep[e.g.,][]{Matteucci86}.
To include SNeIa, we use the analytical prescription of \cite{Greggio05}, which reduces all stellar and binary evolution parameters to the factor $P_\text{SNIa}=6 \cdot 10^{-3}$, representing the probability of an IMS to be born in a system that fulfills all necessary prerequisites to later end up in a SNIa. When the system has reached the end of its lifetime, we eject stable isotopes in the amounts calculated by \citet[][model CDD2]{Iwamoto99}, together with $10^{-4}$\,$\mathrm{M}_\odot$ of \iso{53}Mn, at the same location \citep[which is in agreement with, e.g.,][]{Seitenzahl13,kobayashi20ia}.
When a supernova explosion occurs, a shock wave pushes the ejecta into the ISM; we model this by moving the gas mass of the inner cells into a shell, with an enclosed (pre-explosion) mass of $5\cdot 10^4$ $\mathrm{M}_\odot$. This mass corresponds to an explosion energy of 1 Bethe, according to Sedov-Taylor blast wave theory
\citep{Ryan96,Shigeyama98}, which implies that the radius of a remnant depends strongly on the ISM density surrounding the explosion.
The pushed out gas is distributed in a chemically homogeneous shell around the remnant, leaving behind a ``bubble'' in the ISM. We follow the approach of \cite{Wehmeyer15,Wehmeyer19} and eject a constant yield of elements per SNIa, independent of metallicity. While this approximation is somewhat inaccurate (e.g., \citealt{Timmes03,Thielemann04,Travaglio05,Bravo10,Seitenzahl13,kobayashi20ia}), it does not strongly affect the outcomes of our simulations, which are focused on SLRs that are mostly influenced by solar metallicity yields.
\subsubsection{HMS}\label{Sec:Model:HMS}
HMSs experience every stellar burning stage and produce significant amounts of {$\alpha$}- as well as iron-group elements (e.g., \citealt{Woosley95,Kobayashi06,Limongi18,Ritter18}). When a HMS has reached the end of its life time, we let it explode as a CCSN, analogous to the explosion of a SNIa: We eject stable elements \citep[yields from][]{Thielemann96,Nomoto97} and SLRs (according to Section~\ref{Sec:Model:SLRs}), and move the surrounding gas into a shell around the explosion, depending on the injected kinetic energy.
As described in Section~\ref{Sec:Model:Supernova ejecta dynamics}, we consider a range of different explosion energies and remnant geometries to also account for the potential effect of hypernovae \citep[e.g.,][]{Nomoto04,Nomoto13} and altered CCSN remnant geometries \citep[e.g.,][]{Fry20}. For our purposes, HMS and their CCSNe are the exclusive contributors of the SLR \iso{60}Fe.
\subsubsection{Neutron Star Mergers}\label{Sec:Model:CBM}
Most HMS are born in binary systems (e.g., \citealt{Sana12,Duchene13}), where both stars eventually undergo CCSN explosions, leaving behind two neutron stars (NSs). There is a possibility that these two NSs are still gravitationally bound after the two CCSNe \citep[e.g.,][]{Tauris17}. If in a suitable orbit, these two objects reduce their separation distance via the emission of gravitational waves, until they coalesce. Such a merger event provides conditions to synthesize $r$-process elements \citep[e.g.,][]{Freiburghaus99,Panov08,Korobkin12,Bauswein13,Rosswog13a,Rosswog13b,Rosswog14,Wanajo14,Eichler15,Just15,Vassh19}. For the purpose of our simulation, we reduce all the mentioned probabilities to a factor $P_\text{NSM}$, which represents the fraction of HMSs that fulfil all the needed prerequisites to later undergo a merger event. This approach simplifies the detailed physics of population synthesis, the explosion dynamics of CCSNe, and binary survival probabilities, and allows us to reduce all these details to one free parameter.
We choose $P_\text{NSM}= 0.04$ as in \cite{Wehmeyer19}. Using a Salpeter initial mass function with an integrated slope of $-1.35$, and a standard cosmic star formation history with constant NSM delay times (see \citealt{Cote17} for details), this probability translates to $1.03 \cdot 10^{-4}$ NSM events per unit solar mass of stars formed, which would produce a theoretical gravitational wave event rate of $\sim 1800$ Gpc$^{-3}$ yr$^{-1}$. 
Although this is $\sim 2$ times higher than the latest upper limit of 810 Gpc$^{-3}$ yr$^{-1}$ derived by LIGO/Virgo \citep{2021ApJ...913L...7A}, our main conclusions are not affected by the exact choice of NSM rate (see discussion in Section~\ref{sec:results:dominant prop mechanism}). For our purposes, NSMs are the only source of \iso{244}Pu in the Galaxy.
Analogous to the explosion of a SNIa or a CCSN, when an NSM occurs in our model, \iso{244}Pu is ejected to a spherical shell around the source, assuming an explosion energy of 1 Bethe.

CCSNe mostly occur asymmetrically, resulting in a natal kick for the newly born NS, which could lead to offsets in the NSM locations from the original CCSNe that formed the two NSs \citep[e.g.,][]{vandeVoort22}. However, various studies \citep[e.g.,][]{Beniamini16,Tauris17} have shown that the second born NS in the majority of NS binaries in our Galaxy were formed by much weaker explosions \citep[possibly ultra-stripped SNe, see][]{Tauris15,Mor23} that resulted in very weak kicks. Indeed, \cite{Perets21} showed that the offset locations of short GRBs --- when divided according to galaxy type --- support the idea that kicks play a subdominant role in setting binary neutron star merger offsets. In addition, if binary NS formation is often preceded by such weak explosions, this could lead to the amount of swept up mass by such explosions to be significantly smaller than obtained for standard CCSNe.
\\
Further, the coalescence time of two NSs can be approximated with a $t^{-x}$ distribution \citep[e.g.,][]{Belczynski16,Cote17}, which could also potentially lead to a larger spectrum in abundances \citep[see][]{Cote19b}. Here, we use instead a constant coalescence time of $10^8$ years for newly born NSs to merge.
In this respect, our work will mostly highlight the ``surfing'' effect on SLRs.

\subsubsection{SLR sources and yields}\label{Sec:Model:SLRs}
In this work, we focus on four SLRs: \iso{53}Mn, \iso{60}Fe, \iso{182}Hf, and \iso{244}Pu. For a quick overview over these isotopes' origins in our model, and their half-lives, see Table~\ref{Sec:Model:SLRs:Table}. The deep-sea detections of three of these isotopes (\iso{53}Mn, \iso{60}Fe, and \iso{244}Pu) can be translated into their corresponding ISM density at time of deposition into the deep-sea archive \citep[see][Supplemental Material]{Wallner21}. In our simulation, the four isotopes are produced in the four separate individual sites described above exclusively: SNIa for \iso{53}Mn, CCSNe for \iso{60}Fe, IMSs for \iso{182}Hf, and NSMs for \iso{244}Pu. These associations are reasonable in first order approximation (see details below), furthermore, the advantage of assigning each of these isotopes to a different site is that we are able to use them as unique tracers of each site. Hence we are able to study the production parameters of each site (e.g., occurrence frequency, yields) almost completely independent of the production parameters of the other sites.
We adopted the following yields.
\begin{itemize}
    \item For \iso{53}Mn, $10^{-4}$ $\mathrm{M}_\odot$ is ejected per SNIa. The calculations of \cite{Seitenzahl13} resulted in a range of $3.06 \cdot 10^{-5}$ $\mathrm{M}_\odot$ up to $3.95 \cdot 10^{-4}$ $\mathrm{M}_\odot$ of \iso{53}Mn produced by a SNIa, so our chosen value is well within these limits. It is important to note that the \cite{Seitenzahl13} results were obtained using Chandra-mass models, while our double-degenerate models would probably not produce as much \iso{53}Mn, which would lead to a possible over-estimation of \iso{53}Mn in our model. Although \iso{53}Mn can in principle also be produced by CCSNe \citep[][]{Lugaro16}, the expected yields are rather low and the production is not necessarily efficient around solar metallicities in the galaxy \citep[e.g.,][]{Kobayashi15,Kobayashi20}. Hence, we assume that no \iso{53}Mn is ejected from CCSNe in our model, and test a simplified approach where \iso{53}Mn is ejected by SNIa exclusively.
    \item For \iso{60}Fe from CCSNe, we used progenitor mass-dependent yields from the solar metallicity models by \cite{Limongi06a}. \iso{60}Fe might also be produced in electron-capture supernovae \citep[e.g.,][]{Wanajo13}; at solar metallicity, however, the predicted rate for this site is expected to be low ($\sim 1-5$\%) in comparison to CCSNe \citep[e.g.,][]{poelarends08,doherty15,Kobayashi20}, and the ejecta mass is $\sim$ 100 times lower than that of CCSNe. Therefore, we do not consider it as a source of \iso{60}Fe in our model.
    \item For \iso{182}Hf from IMSs, we used progenitor mass-dependent yields from the solar metallicity models by \cite{Lugaro14}. \iso{182}Hf is also produced in NSMs, but since their occurrence frequency is much lower than IMSs, they are far less relevant for the overall production of \iso{182}Hf, and we thus simplify the model to produce \iso{182}Hf only in IMSs. Further, this simplification allows us to test whether \iso{182}Hf would still arrive in deep-sea sediments conjointly with the CCSN produced \iso{60}Fe and the NSM produced \iso{244}Pu, even though it is not produced by either of them in our model.
    \item The yields for \iso{244}Pu from NSMs are highly uncertain. If we assume the order of $10^{-2}$ $\mathrm{M}_\odot$ total mass ejection by a single NSM \citep[e.g.,][]{Korobkin12}, and a mass fraction of $10^{-6}$ for \iso{244}Pu (which is within the range predicted by \citealt{Eichler15} using different fission fragment and nuclear mass models), we obtain $10^{-8}$ $\mathrm{M}_\odot$ \iso{244}Pu ejected by a single NSM, the value used for our simulations. Black hole -- neutron star mergers might also produce \iso{244}Pu if the mass of the black hole is low enough, or its spin is sufficiently strong, however, this site is probably more relevant at early galactic stages. At later galactic stages, the occurrence rate of NSMs likely exceeds the occurrence rate of black hole -- neutron star mergers \citep[e.g.,][]{Wehmeyer19}. This is consistent with population synthesis models which typically predict that NSMs occur more frequently than black hole -- neutron star mergers at high metallicities \citep[e.g.,][]{Dominik12,Chruslinska19}. Hence, in our simulations, we assume that \iso{244}Pu is dominated by NSMs.
In this work, we omit also other possible $r$-process sites such as magnetorotationally driven CCSNe \citep[e.g.,][]{Winteler12,Nishimura17,2018ApJ...864..171M,2021MNRAS.501.5733R} and rare, peculiar CCSNe, e.g., hadron-quark phase transition CCSNe \citep{Fischer20}, as these sites still lack observational confirmation.
\end{itemize}
We comment on the dependence of the results on CCSN, IMS, and NSM yields in Section~\ref{sec:results:evolutionof  SLRS:Yields}.

\begin{table}
\centering
\begin{tabular}{|l|c|l|}
\hline
\hline
Isotope &Half-life t$_{1/2}$ & Source\\
\hline
\iso{53}Mn & 3.74 Myr& SNIa\\
\iso{60}Fe & 2.62 Myr& CCSN\\
\iso{182}Hf & 8.90 Myr& IMS\\
\iso{244}Pu & 80.0 Myr& NSM\\
\hline
\end{tabular}
\caption{Overview of the isotopes, their half-lives, and their sources in our model. See Section~\ref{Sec:Model:SLRs} for details.}
\label{Sec:Model:SLRs:Table}
\end{table}

\subsubsection{Supernova ejecta dynamics}\label{Sec:Model:Supernova ejecta dynamics}
To estimate the influence of hypernovae, featuring substantially higher explosion energies than regular CCSNe \citep[e.g.,][]{Nomoto04,Nomoto13}, and the effect of a varied CCSN bubble remnant geometry and SLR distribution within the explosion shell due to magneto-hydrodynamical effects \citep[e.g.,][]{Fry20}, we set up four different scenarios (Table~\ref{Analysis overview:table}) to study the implications of different assumptions for the interaction of CCSN ejecta with the ISM.
\begin{enumerate}
    \item \textit{Standard case} (as described in Section~\ref{Sec:Model:HMS}). All CCSNe explode with a kinetic energy of 1 Bethe, and therefore pollute $5\cdot 10^4 $ $\mathrm{M}_\odot$ of ISM.  We assume that all SLRs are deposited on a chemically well mixed shell located on the edge of the blast wave.
    \item \textit{Increased explosion energy case (hypernova model, HN).} All CCSNe have an increased explosion energy and pollute $2\cdot 10^5 $ $\mathrm{M}_\odot$ of ISM, this estimates the effect on the SLR abundance evolution if all CCSNe exploded as hypernovae. The remnant geometry is the same as in the standard case.
    \item \textit{Modified geometry case (PINBALL).} To estimate the impact of a potential ``pinball model''-style remnant geometry \citep{Fry20}, where SLRs are reflected backwards towards the center of the explosion after the remnant bubble has halted, all CCSNe explode with a kinetic energy of 1 Bethe, but we assume that $1\%$ of the swept-up ISM (and therefore SLRs) contained therein will remain homogeneously
    distributed inside the explosion bubble.
    \item \textit{Combination of increased explosion energy and modified geometry (HN PINBALL).} As a combination of models HN and PINBALL, all CCSNe pollute $2\cdot 10^5 $ $\mathrm{M}_\odot$ of ISM and $1\%$ of the swept-up ISM are distributed homogeneously inside the explosion bubble.
\end{enumerate}
Once these values/scenarios are chosen at the beginning of a simulation, they remain constant throughout the entire run. This is a simplification because 
supernova (SN) explosion radii are sound speed dependent, and thus also dependent on the local density \citep[e.g.,][]{Chamandy20}, which   in a galaxy strongly fluctuates, especially in its earlier evolution stages. 
However, for our modelling, we use a Sedov-Taylor approach, which simplifies these assumptions to an almost constant swept-up mass mostly determined by the CCSN explosion energy \citep[e.g.,][who employed a constant sound speed of 10km/s]{Shigeyama98}, this swept-up mass parameter remains constant during a run once it is chosen in the beginning. This approach might underestimate the spectrum of the SLR densities, especially at earlier galactic stages, but we chose to keep this swept-up mass parameter constant to be able to focus on identifying the dependence of SLR density on mixing caused by explosions. Further, our approach limits the time resolution of the model. We deliberately chose a time step size of 1 Myr because this allows us to simplify all thermo- and hydrodynamic processes into the value of the swept-up mass. One Myr after explosion, the shock front has stopped, and the ejected and swept-up material has cooled down sufficiently to be available for star formation in the following time step. With better time resolution (with shorter time steps), the explosion ejecta and the swept-up material would have to be closely followed, which would require a detailed thermo- and hydro-dynamic modeling, not available in our model. We tested, instead, the dependence of our model on a decreased rather than increased time resolution (i.e., longer time steps) in Appendix~\ref{appendix:sec:resolution}.
\begin{table}
\centering
\begin{tabular}{|l|l|l|}
\hline
\hline
Model name & ISM polluted & Remnant geometry\\
\hline
Standard & $5\cdot 10^4$ $\mathrm{M}_\odot$ & Standard\\
HN & $2\cdot 10^5 $ $\mathrm{M}_\odot$ & Standard\\
PINBALL & $5\cdot 10^4 $ $\mathrm{M}_\odot$ & PINBALL model\\
HN PINBALL & $2\cdot 10^5 $ $\mathrm{M}_\odot$ & PINBALL model\\
\hline
\end{tabular}
\caption{Overview of the different models. The left column states the model name, the middle column how much ISM is polluted by a single CCSN, and the third column how the swept-up material is distributed after the event. See Section~\ref{Sec:Model:Supernova ejecta dynamics} for details.}
\label{Analysis overview:table}
\end{table}
\section{Results} \label{sec:results}
In this section, we report on the evolution of SLRs over the lifetime of the Galaxy (Section~\ref{sec:Results:Evolution of SLRs}), and compare our results to deep-sea sediment abundances (Section~\ref{sec:results:recentevolution}).
\subsection{Evolution of SLRs}\label{sec:Results:Evolution of SLRs}
\begin{figure*}
\centering
\subfigure[Standard model]{\includegraphics[width=\columnwidth]{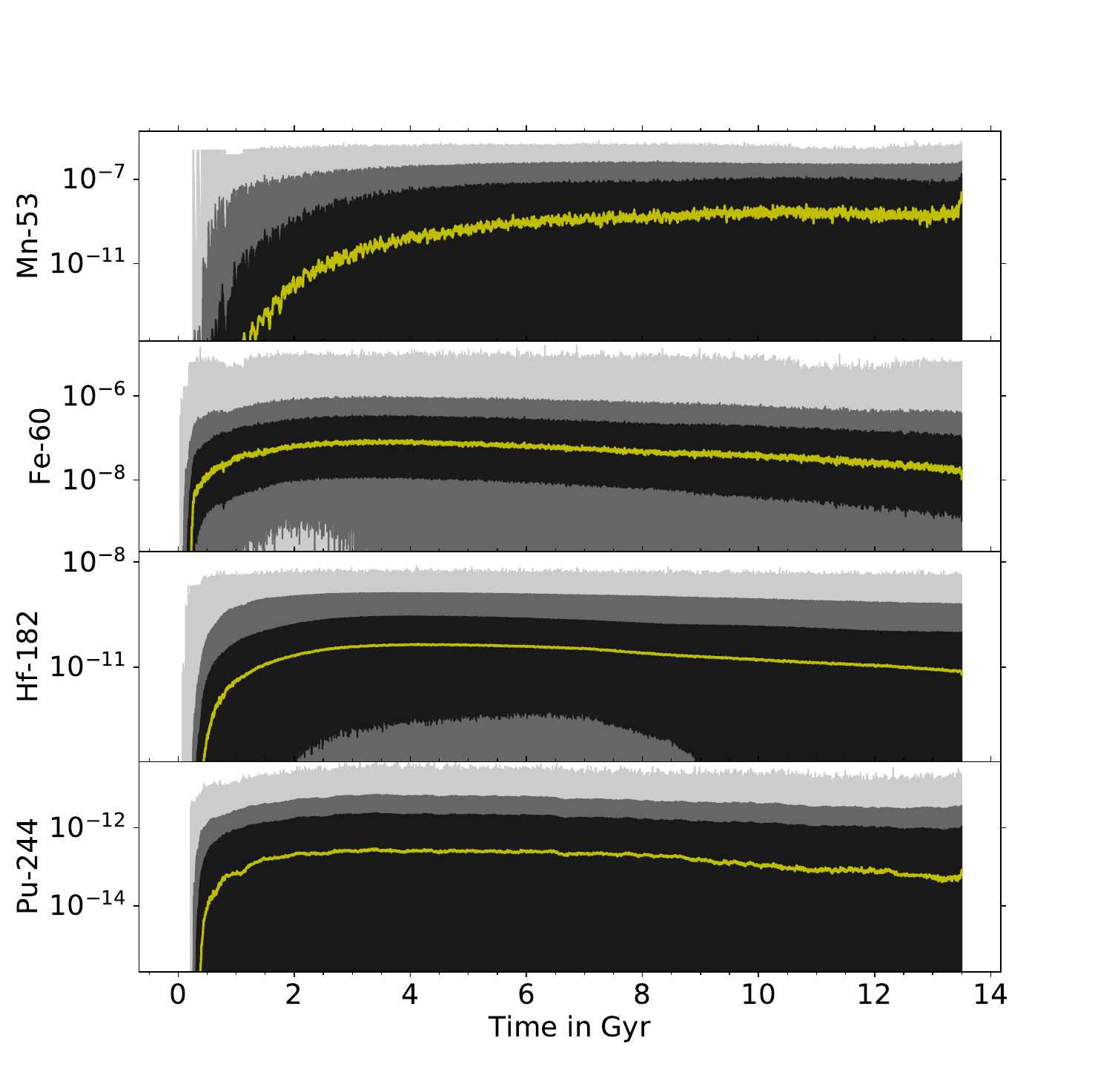}} \hfill
\subfigure[HN model]{\includegraphics[width=\columnwidth]{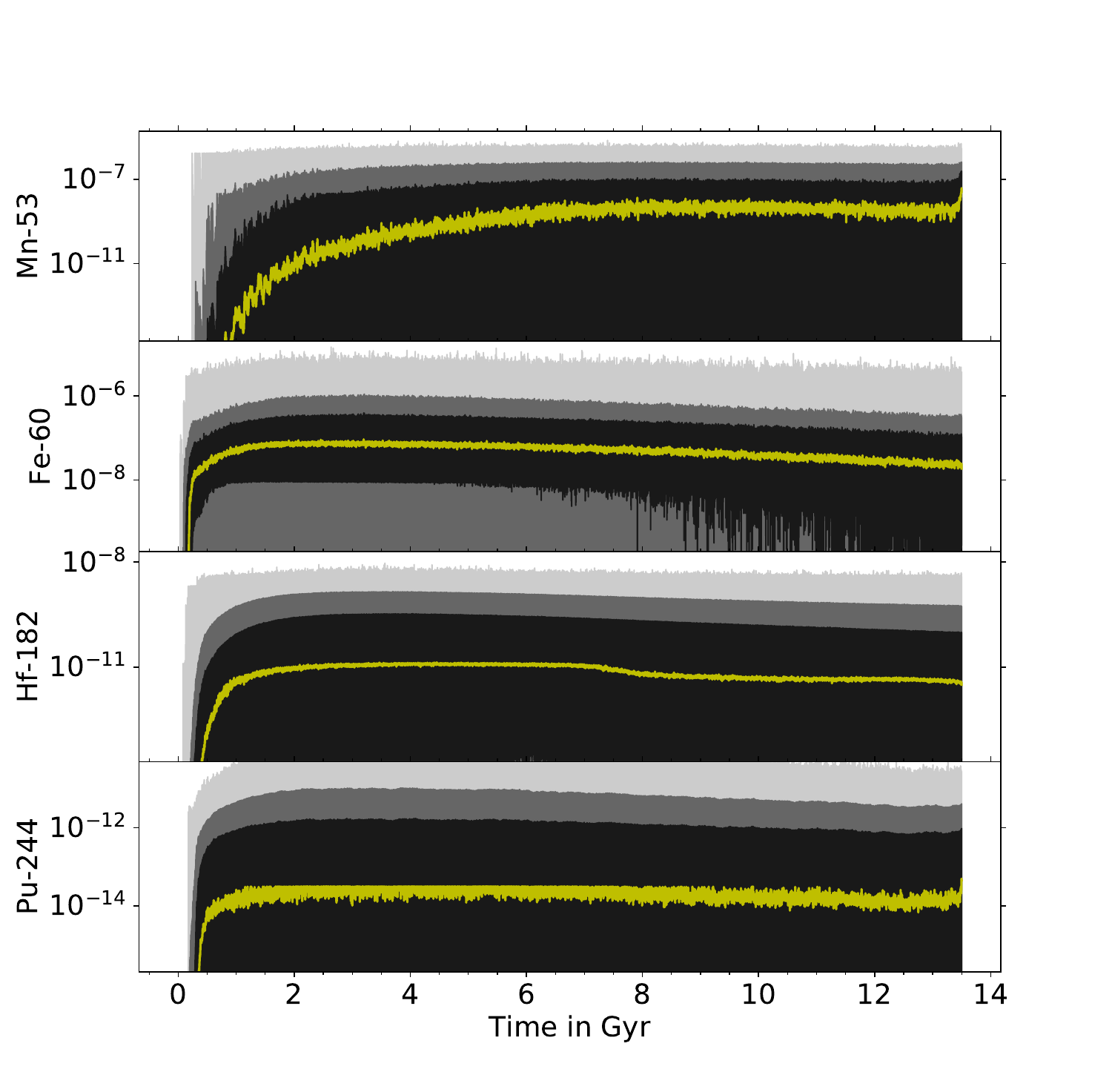}}
\subfigure[PINBALL model]{\includegraphics[width=\columnwidth]{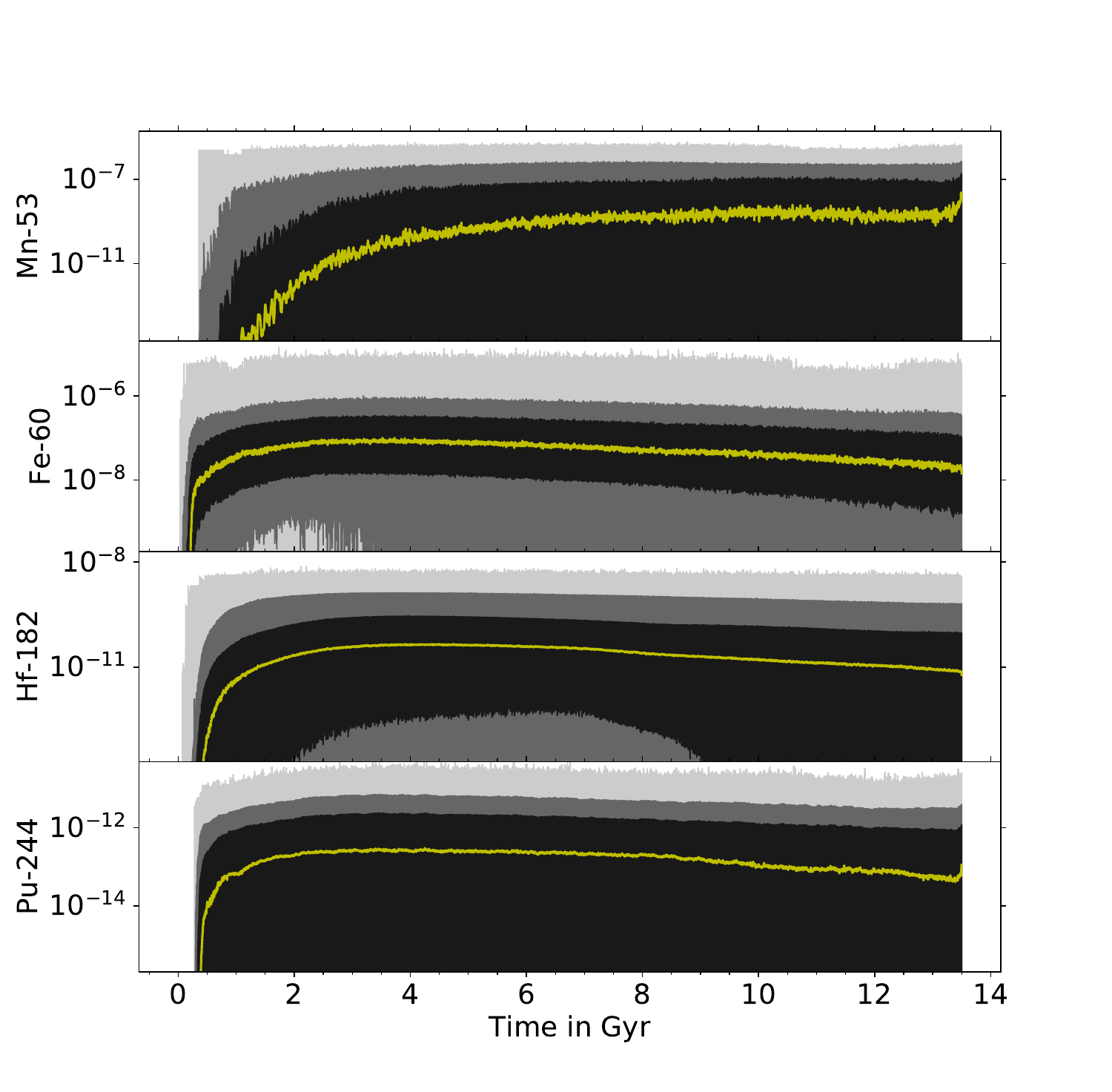}}\hfill
\subfigure[HN PINBALL model]{\includegraphics[width=\columnwidth]{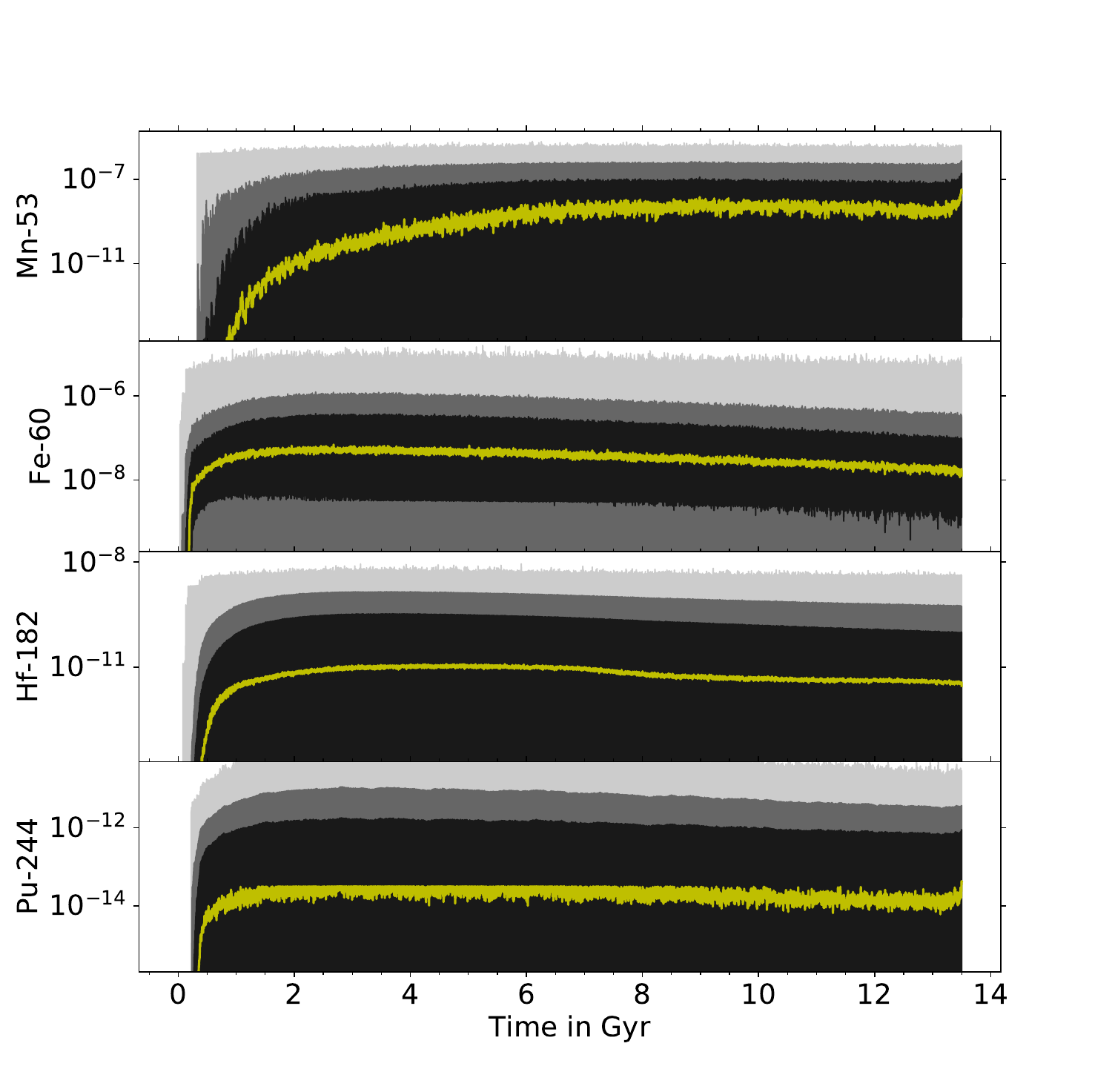}}
\caption{SLR densities in g/cm$^3$ in the simulation volume, where the distribution at each time step represents the spatial distribution of SLR densities in the (40)$^3$ sub-cells, for the four cases of Table~\ref{Sec:Model:SLRs:Table}, as indicated below each panel. The median is shown as a solid yellow line. The black (dark grey, light grey) shaded areas represent the 68\% (95\%, 100\%) distribution.}
\label{Figure:Gas}
\end{figure*}

At every time step, we store and analyse the gas contents and the SLR abundances in every cell. Figure~\ref{Figure:Gas} shows the statistical distribution of the abundances of the four isotopes over the lifetime of the Galaxy for the four cases described in Section~\ref{Sec:Model:Supernova ejecta dynamics}.
Overall, the ISM densities of the four SLRs follow the galactic star formation rate with an early rise and slow decrease, although each of them shows a different shift to later times relative to the peak of the star formation rate. The magnitude of these shifts  (where we define $t_{1/100}^\mathrm{SLR}$ as the delay between the onset of the simulation and the time when the median of the abundance of a given SLR reaches 1\% of its maximum within the first 3 Gyr) is linked to the delay time between the formation of the progenitor stars and the enrichment events. We have summarized the values of $t_{1/100}^\mathrm{SLR}$ for each model in Table~\ref{evolutionofslrs:table}.\\
\begin{table}
\centering
\begin{tabular}{|l|l|l|l|l|}
\hline
\hline
Model name & $t_{1/100}^\mathrm{^{53}Mn}$ &  $t_{1/100}^\mathrm{^{60}Fe}$ & $t_{1/100}^\mathrm{^{182}Hf}$ & $t_{1/100}^\mathrm{^{244}Pu}$\\
\hline
Standard & 1790 & 254 & 585 & 432\\
HN & 138 & 192 &  480 & 370\\
PINBALL & 192 & 228 &  580 & 426\\
HN PINBALL & 152  & 194 &  483 & 360\\
\hline
\end{tabular}
\caption{Delay time ($t_{1/100}^\mathrm{SLR}$, in Myr) for each SLR between the onset of the simulation and the time when the median of the abundance of a given SLR reaches 1\% of its maximum within the first 3 Gyr.}
\label{evolutionofslrs:table}
\end{table}

In our simulations, CCSNe (ejecting \iso{60}Fe) have the shortest delay times, because they are produced from HMSs with the shortest life times. This leads to the \iso{60}Fe curve \ having the smallest shift towards later times ($t_{1/100}^\mathrm{^{60}Fe}=254$ Myr). \iso{244}Pu has the second lowest delay time($t_{1/100}^\mathrm{^{244}Pu}=432$ Myr), since this SLR is ejected as soon as two short-living HMSs have died and the two resulting NSs have spiraled inwards towards their common center of mass for the coalescence time. \iso{182}Hf has a longer delay time than \iso{244}Pu  ($t_{1/100}^\mathrm{^{182}Hf}=585$ Myr), because IMSs stars live longer than HMSs plus our assumed coalescence time. Since we require two IMSs in a binary system to have reached the end of their lives for a SNIa to occur, the time scale to produce the SNIa ejecta will always be longer than the lifetime of the secondary star in the binary system. Thus, \iso{53}Mn has the largest shift of all SLRs ($t_{1/100}^\mathrm{^{53}Mn}=1.79$ Gy).
\\
Because of the large variations of gas contents and SLR abundances among the cells, we present the SLR densities in the cells for every time step as statistical distributions, where the shaded areas represent 100\% (light grey), 95\% (dark grey) and 68\% (black) of the SLR densities. For \iso{60}Fe, the  68\% cell-to-cell fluctuation span two orders of magnitude at almost all times.
The spectrum of possible \iso{60}Fe densities is the most narrow right after the time of highest star formation (at the point of highest number of HMS deaths) at t$\approx 3.5$ Gy, and broadens with later times. The reason why many stellar deaths lead to a narrower spectrum in abundances stems from the fact that the SLR has less time to decay between enrichment events, therefore the minimum abundance value immediately before a subsequent nucleosynthesis event is closer to the maximum value of that abundance just after the nucleosynthesis event, as compared to a case when stellar deaths are more apart in time. In this latter case, the SLR has much more time to decay before the subsequent injection of that SLR into the ISM, and hence decays to lower values than in the former case. A secondary effect is that if more stars die in a given time span, there is also more production of \iso{60}Fe, which leads to a higher overall abundance of \iso{60}Fe. A corresponding behavior can also be observed for the other SLRs, at the respective points in time when the highest number of stars contribute to each SLR.
\\
We show the implications of the four shock remnant models introduced in Section~\ref{Sec:Model:Supernova ejecta dynamics} in Figure~\ref{Figure:Gas}:   Stronger explosions (HN model) lead to a larger spectrum of SLR density. This effect can be well observed when comparing the difference in e.g., the 68\% band of \iso{60}Fe between the standard and the HN model. In the latter, more cells are affected by a single SN explosion. Because of the larger radius of the remnant in the HN model, there are more cells inside the remnant, and these are cleared of their SLR content, which leads to a larger number of lower SLR density cells in every time step. For \iso{244}Pu, this also leads to a significantly lower density median, since NSMs are very rare, the additional number of cells with low \iso{244}Pu content strongly affects the median.
In the HN model, instead, more cells are polluted per time step and thus the mixing is more efficient. Therefore, also the delay for the SLRs to approach their steady-state value is lower than in the Standard model ($t_{1/100}^\mathrm{^{53}Mn}=138$ Myr, $t_{1/100}^\mathrm{^{60}Fe}=192$ Myr, $t_{1/100}^\mathrm{^{182}Hf}=480$ Myr, $t_{1/100}^\mathrm{^{244}Pu}=370$ Myr).
\\
The PINBALL model case leads to ejecta distributed more homogeneously inside the supernova bubble, however, the effect appears to influence the SLR density statistics only marginally.
The delays of the SLRs to approach their steady-state value  
($t_{1/100}^\mathrm{^{53}Mn}=192$ Myr, $t_{1/100}^\mathrm{^{60}Fe}=228$ Myr, $t_{1/100}^\mathrm{^{182}Hf}=580$ Myr, $t_{1/100}^\mathrm{^{244}Pu}=426$ Myr) are all longer than in the HN model case, but show only a slight correction when compared to the Standard model.
Even when the fraction of reflected material is increased to a much higher value (50 \%), it only affects SLR densities locally in sub-cells, but not their overall abundance statistics, as described in Appendix~\ref{appendix:sec:more geo}.
\\
In the combined case with both PINBALL geometry and high explosion energy (HN PINBALL model), instead the flattening effect of the more homogeneous distribution of SLRs (due to the PINBALL treatment) combined with the larger explosion bubbles (due to the HN treatment) becomes more prominent. In this model, ejecta from every nucleosynthesis site are distributed throughout the entire, larger explosion bubble, which results in more cells being polluted by an SLR per nucleosynthesis event than in any other model. For the more rare sites (SNeIa and NSMs producing \iso{53}Mn and \iso{244}Pu, respectively), this also leads to larger fluctuations in the median between time steps (compared to the standard model), since the SLRs produced in these rare events are distributed throughout the volume much quicker. This behavior can also be observed in the delay times for the SLRs to approach the steady-state value ($t_{1/100}^\mathrm{^{53}Mn}=152$ Myr, $t_{1/100}^\mathrm{^{60}Fe}=194$ Myr, $t_{1/100}^\mathrm{^{182}Hf}=483$ Myr, $t_{1/100}^\mathrm{^{244}Pu}=360$ Myr). This means that the ISM homogenizes faster, eliminating quickly the spikes in SLR densities generated by a single rare nucleosynthesis event. In other words, since this last model homogenizes the ISM so quickly, the median behaves more like a one-zone-model instead of a single cell in the simulation volume, the latter will be discussed in the following section.
\subsection{Recent evolution}\label{sec:results:recentevolution}
\begin{figure*}
    \centering
\subfigure[Standard model]{\includegraphics[width=\columnwidth]{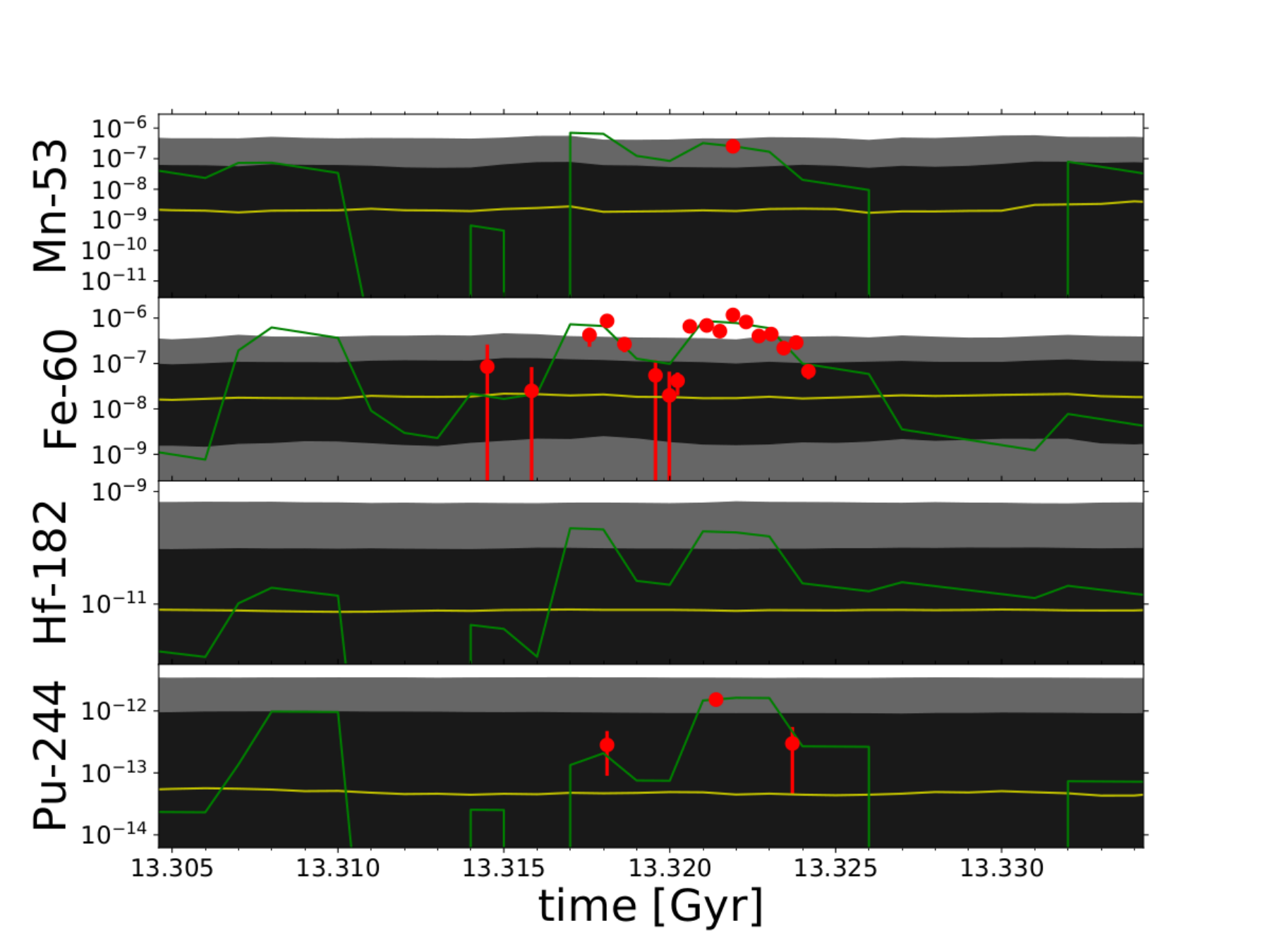}}\hfill
\subfigure[HN model]{\includegraphics[width=\columnwidth]{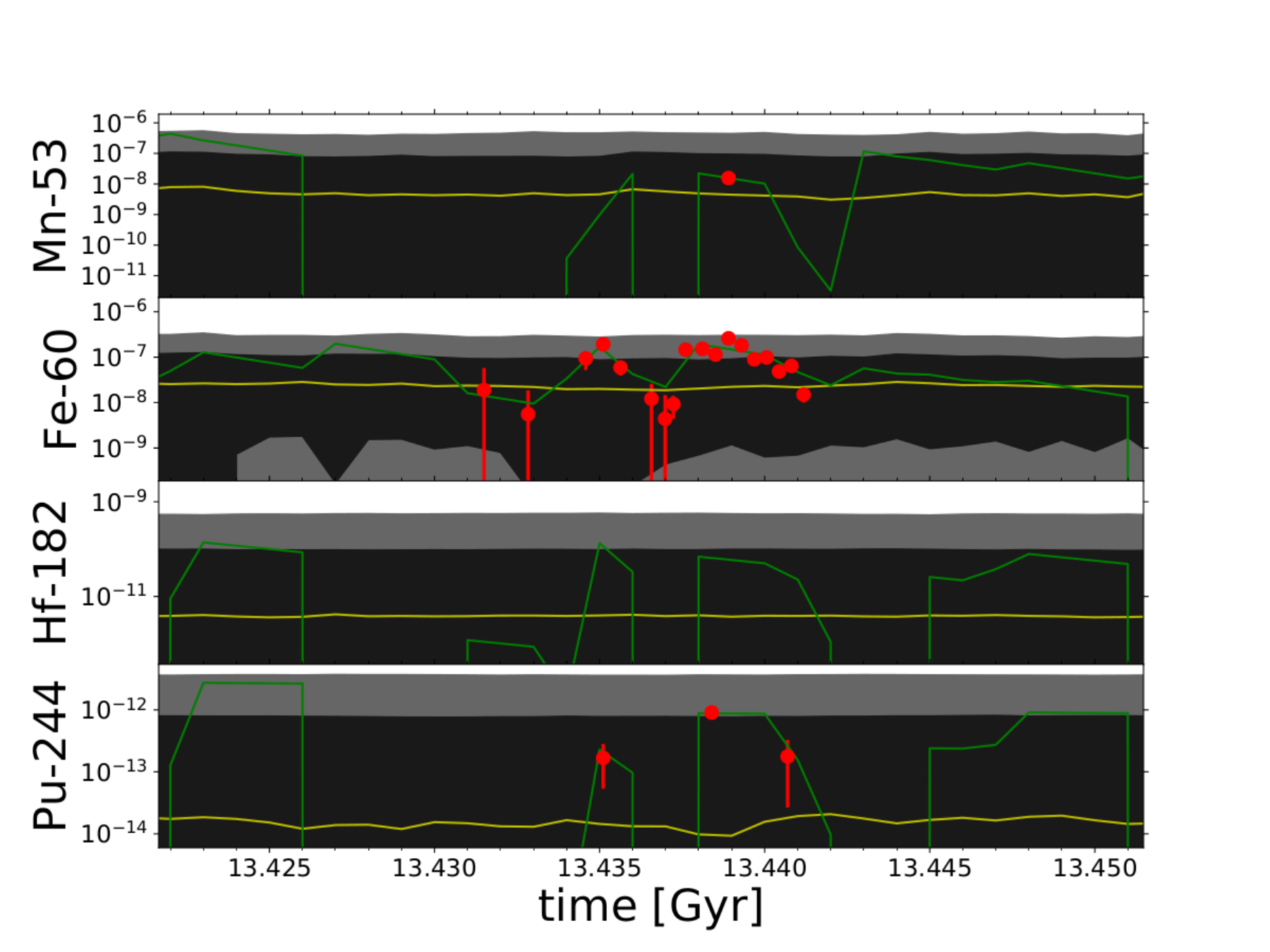}}
\subfigure[PINBALL model]{\includegraphics[width=\columnwidth]{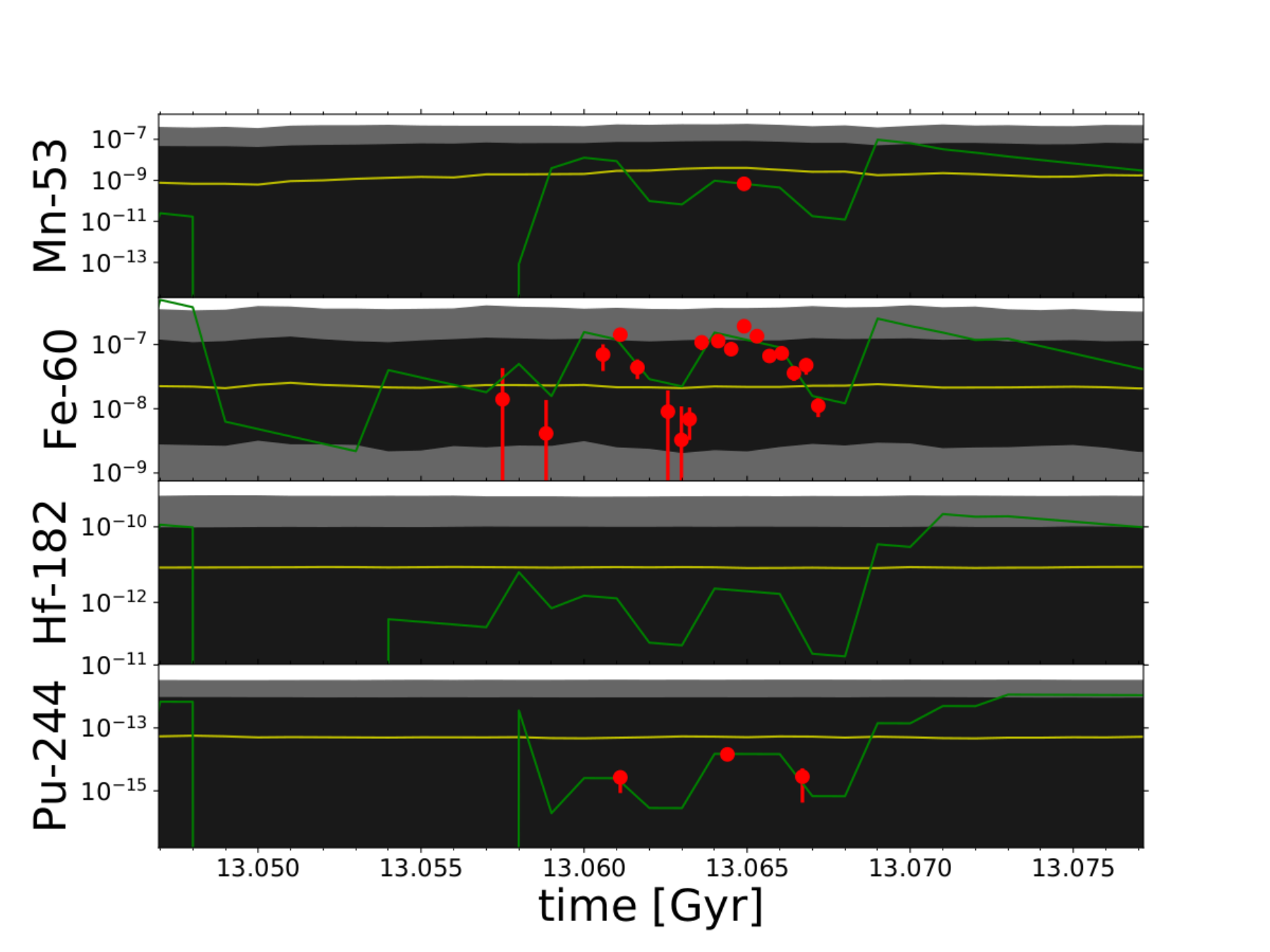}}\hfill
\subfigure[HN PINBALL model]{\includegraphics[width=\columnwidth]{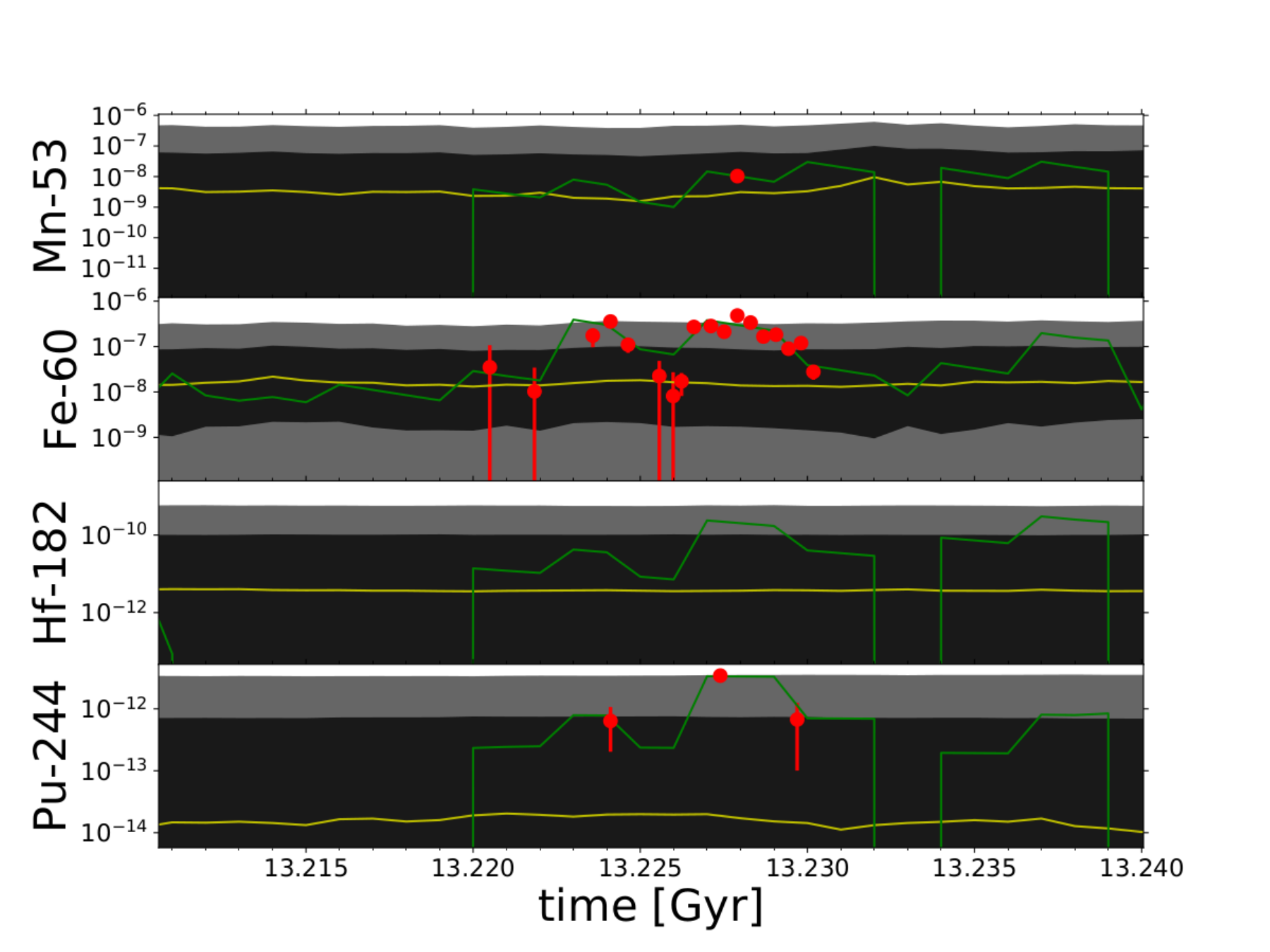}}
\caption{Same as Figure 1, but zoomed-in to $\pm$10 Myr around the respective $\Delta$t. In addition to what is shown in Figure 1, ISM densities for \iso{53}Mn, \iso{60}Fe, and \iso{244}Pu, inferred from deep-sea sediment detections are shown as red symbols with error bars (shifted by factors $\lambda$ and $\Delta$t), and the abundance evolution of the one simulation cell in each model that best fits these deep-sea detection ISM abundances is shown in green.} 
\label{fig:recentevolution}
\end{figure*}

Figure~\ref{fig:recentevolution} shows a zoom-in on the evolution of the SLRs, closer to the current day. To the figure, we added the ISM densities inferred from deep-sea sediment detections from \cite{Wallner16}, \cite{Korschinek20} and \cite{Wallner21}, as well as the SLR density evolution in one of the sub-cells of the simulation which best fits the ISM densities derived from the deep-sea detections. We introduced a factor $\lambda$ for a vertical shift for all derived ISM densities of every isotope of the deep-sea detections in every model for fitting, since we were more interested in fitting the shape of the detection curves, rather than the actual values. Further, we introduced a factor $\Delta$t to account for a horizontal (time) shift in the ISM densities. The values for the two factors for each model can be found in Table~\ref{Recentevolution:table}.
We test the possibility for a sub-cell to fit the detection data in Appendix~\ref{appendix:sec:robustness}.
\begin{table}
\centering
\begin{tabular}{|l|l|l|l|l|}
\hline
\hline
Model name & $\lambda_{53_{\mathrm{Mn}}}$ &  $\lambda_{60_{\mathrm{Fe}}}$ & $\lambda_{244_{\mathrm{Pu}}}$ & $\Delta$t (Myr)\\
\hline
Standard & 4.07 & 0.211 & 16.5 & 174.63\\
HN & 4.15 & 0.125 &  1.01 & 57.63\\
PINBALL & 4.24 & 0.02 &  0.0439 & 431.63\\
HN PINBALL & 4.23 & 0.476 &  0.669 & 240.63\\
\hline
\end{tabular}
\caption{Vertical ($\lambda$) and horizontal (time) shift ($\Delta$t) factors for the vertical and the horizontal shift of the ISM densities of the four isotopes as inferred by their deep-sea detections used for our fitting.}
\label{Recentevolution:table}
\end{table}
\\
First, we consider the top left panel in Figure~\ref{fig:recentevolution} (the standard case). If we follow the density evolution of the best-fitting cell, the effects of radioactive decay of the SLRs in that cell are visible (e.g., around $13330$ Myr in the \iso{60}Fe evolution).
However, another effect often dominates: since we assume that SN ejecta behave in a Sedov-Taylor-like expansion pattern, any stellar explosion clears its neighboring cells completely of their contents as the blast wave travels through those cells. Together with the gas contents, also the SLRs are carried away from those cells to pile up in the SN remnant shell. This results in a very low or even zero gas and isotope content of these cells, which is the main cause of the major discontinuities seen in the green lines in the figure. 
\\
Further, the sudden increases in the green lines can also be explained by blast waves. Again, all cells within the radius of a shock wave are emptied (or almost emptied in the PINBALL models), and their gas and isotope content from before the explosion is deposited on a shell around the explosion. If an observed cell is located on the shell around an explosion, its gas and isotope content thus strongly increases, which leads to an upward jump in isotope abundances in the cell.
\\
This effect of blast waves is observed in all SLRs in our models and might explain the data of \cite{Wallner21}, who found that \iso{60}Fe and \iso{244}Pu arrive on Earth synchronously. If in our model \iso{60}Fe suddenly arrives in an ISM cell due to a nearby stellar explosion, it is very likely that also some \iso{244}Pu arrives synchronously in that cell.
\\
In the top right panel of Figure~\ref{fig:recentevolution} (increased explosion energy), the sweep-up effect discussed above is more frequent as demonstrated by the fact that the lower 68\% confidence band has a lower boundary), which happens because more space (i.e., more simulation cells) is affected by a each CCSN, therefore, the clearing out effect occurs more frequently than in the standard case.
For the PINBALL model case (bottom left panel of Figure~\ref{fig:recentevolution}), we observe the same as in the previous Section, that this model does not have a strong effect on the statistics of the SLR densities. A difference to the standard model, however, can be observed here when considering the evolution in a single cell (green line), which is more variable than in the standard model case (when the radioactive decay effect is subtracted from the evolution of the line). This is because more cells are affected per CCSN explosion. This leads to all cells being affected more often in a given time interval, which leads to the density in each cell varying at a higher frequency as compared to the standard case.
\\
In the lower right panel of Figure~\ref{fig:recentevolution} (HN PINBALL model), the evolution of the single cell (green line) oscillates even stronger in comparison to all the other models, because the most cells are affected by each single CCSN. This results in any given cell being affected by SLR density changes much more often in a given time span, which results in the density curve oscillating at the highest frequency. In this model, the SLR density in a given cell is completely determined by external events, rather than by radioactive decay, whereas decay is the dominant effect in the standard model.
\\
In all four models, it is possible to find cells that reproduce the shape of the measurement data reasonably well.
The underlying reason for the observation that all four isotopes show a synchronous increase in a given cell is that CCSNe are the dominant propagation mechanism for all SLRs. Even if e.g., an NSM (ejecting \iso{244}Pu) happened far away from a given location in the ISM, CCSNe will be responsible for the ejecta propagation, as we will see in the following Sections.
\section{Discussion}
In this section, we analyze the pollution intervals to a given parcel of the ISM expected for different SLRs (Section~\ref{sec:results:Pollution intervals}), and the propagation of SLRs in the ISM (Section~\ref{sec:results:dominant prop mechanism}). We suggest a schematic interpretation of the findings of Sections~\ref{sec:results:Pollution intervals} and \ref{sec:results:dominant prop mechanism} in Section~\ref{sec:schematic_interpretation}.
Further, we test the impact of varying yields and event occurrence frequencies on our results in Section~\ref{sec:results:evolutionof SLRS:Yields}.
\subsection{Pollution intervals}\label{sec:results:Pollution intervals}
\begin{figure}
    \centering
    \includegraphics[width=0.99\linewidth]{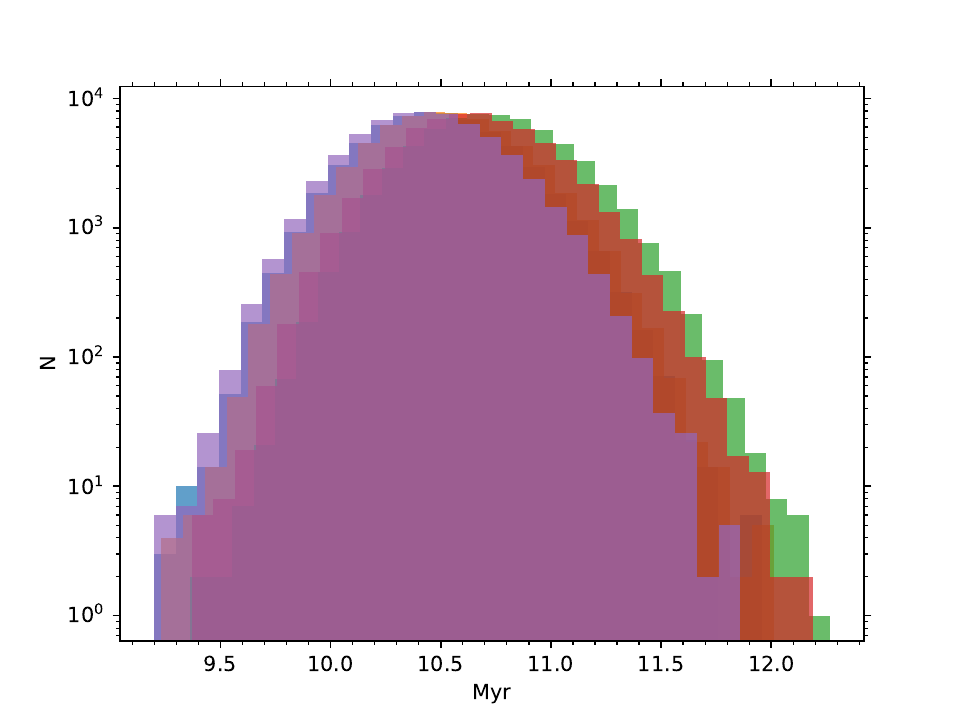}
    \caption{Distribution of the means of the elapsed time of arrival of the different focus SLRs in all cells: $<\delta_{\mathrm{^{53}Mn}}>$ (red), $<\delta_{\mathrm{^{60}Fe}}>$ (yellow), $<\delta_{\mathrm{^{182}Hf}}>$ (green),  $<\delta_{\mathrm{^{244}Pu}}>$ (magenta), and $<\delta_\mathrm{mass}>$ (blue).}
    \label{fig:Pollution_intervals}
\end{figure}

We have noted in the previous section that the increases in different SLR densities often coincide (green lines in Figure~\ref{fig:recentevolution}), although they are produced by different nucleosynthesis sites with very distinct occurrence frequencies. To investigate the cause of this behavior, we examine how often a given gas cell is polluted with a given SLR (or equivalently, how much time elapses between two consecutive pollution events in that cell), and compare this to the time that elapses between two consecutive positive changes in the gas mass of the cell\footnote{We examine this for the standard model only, although the conclusions drawn could be extended to the other models as well.}. To exclude infalling gas (see Section~\ref{Sec:Model:General setup}) from triggering a positive gas mass change event in a cell, we set a fiducial threshold for such a positive change in gas to 100 solar masses of gas or more (which is higher than the highest amount of infalling gas at all times).
We monitored the time between two consecutive SLR and gas mass enrichment events, $\delta_\mathrm{SLR}$ (i.e., $\delta_\mathrm{^{53}Mn}$,  $\delta_\mathrm{^{60}Fe}$, $\delta_\mathrm{^{182}Hf}$, or $\delta_\mathrm{^{244}Pu}$), and $\delta_\mathrm{mass}$, respectively, in every cell. Then, we calculate the average of each of these pollution times ( $< \delta_\mathrm{SLR} >$ for every SLR) in each cell over the entire lifetime of the Galaxy, and generate a histogram over the averages of all cells, which is shown in Figure~\ref{fig:Pollution_intervals}. All the histogram curves show a Gaussian behavior (due to the central limit theorem) and overlap with each other at large portions. This means that all the SLRs arrive at a given location at a very similar frequency, and in conjunction with every arrival of gas. This is further evidence that the probability of all SLRs to arrive synchronously at a given location is very high.
\subsection{The dominant propagation mechanism}\label{sec:results:dominant prop mechanism}
\begin{figure}
    \centering
    \includegraphics[width=0.99\linewidth]{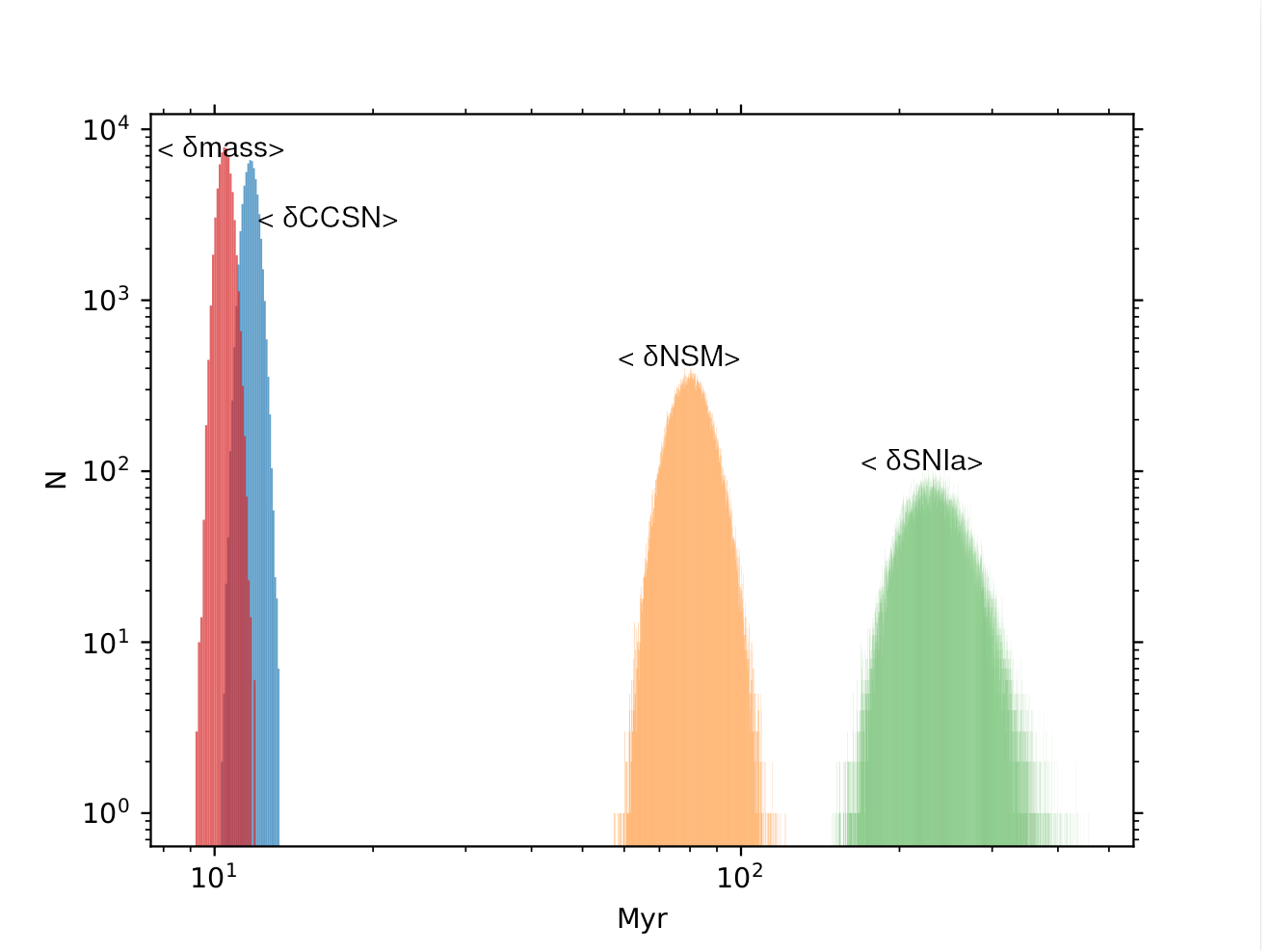}
    \caption{Cell distribution of the means of the elapsed time between two consecutive events of the same kind: $<\delta_\mathrm{CCSN}>$ (blue), $<\delta_\mathrm{NSM}>$ (yellow), $<\delta_\mathrm{SNIa}>$ (green), and $<\delta_\mathrm{mass}>$ (magenta, same as blue color curve in Figure~\ref{fig:Pollution_intervals}).}
    \label{fig:dominant_mechanism}
\end{figure}

To further examine the cause of the synchronous arrival of SLRs, we explore the contributions of the different nucleosynthesis sites towards the propagation of matter throughout our simulation volume and the age of the Galaxy. For every nucleosynthesis event, we monitor the coordinates of all the cells that are affected. From this, we calculate the time that elapses between two consecutive events that affect each given cell for each type of explosion, $ \delta_\mathrm{site} $ (i.e., $ \delta_\mathrm{CCSN} $, $ \delta_\mathrm{NSM} $, and $ \delta_\mathrm{SNIa} $). In other words, we examine the time that elapses between the  cell being affected by the shock fronts of two different explosions of the same type of event.
We then calculate the mean of these elapsed times for each exploding nucleosynthesis site, $<\delta_\mathrm{CCSN}>$, $<\delta_\mathrm{NSM}>$, and $<\delta_\mathrm{SNIa}>$, and plot these medians in Figure~\ref{fig:dominant_mechanism}, together with $<\delta_\mathrm{mass}>$ from Figure~\ref{fig:Pollution_intervals}. None of the $<\delta_\mathrm{site}>$ distributions is congruent with $<\delta_\mathrm{mass}>$, but $< \delta_\mathrm{CCSN} >$ is the closest to $<\delta_\mathrm{mass}>$, which means that CCSNe affect the most cells, relative to the other types of explosions.
We therefore conclude that CCSNe are the dominant propagation mechanism of mass. When considering that $<\delta_\mathrm{mass}> \approx <\delta_\mathrm{SLR}>$
from Section~\ref{sec:results:Pollution intervals}, we further conclude that CCSNe are the dominant propagation mechanism of all SLRs in our model.
This means that the ejecta from all the nucleosynthesis sites included in our model travel as dynamically as the ejecta of CCSNe, even if their occurrence rate is much lower than that of CCSNe.
\subsection{Schematic interpretation} \label{sec:schematic_interpretation}
\begin{figure}
    \centering
 \begin{minipage}[t]{0.5\textwidth}
 \centering
    \includegraphics[width=0.6\linewidth]{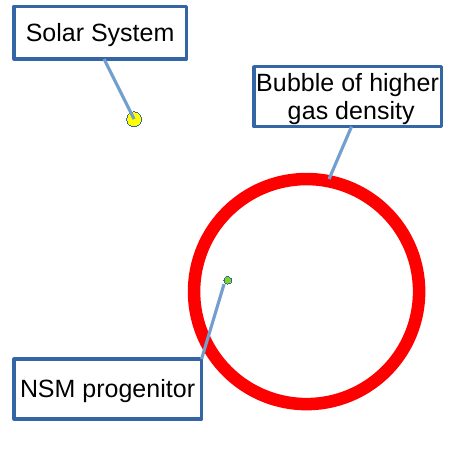}
    \end{minipage}
 \begin{minipage}[t]{0.5\textwidth}
 \centering
        \includegraphics[width=0.6\linewidth]{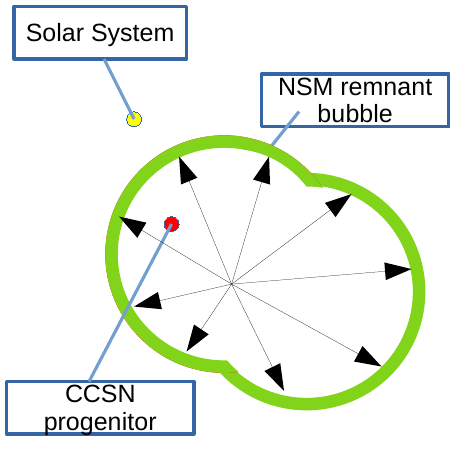}
                \end{minipage}
 \begin{minipage}[t]{0.5\textwidth}
 \centering
\includegraphics[width=0.6\linewidth]{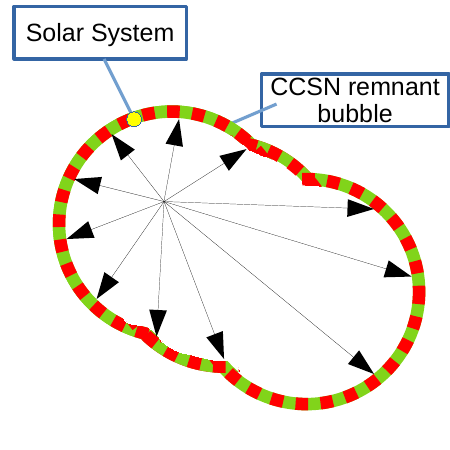}
\end{minipage}
    \caption{Schematic interpretation of the the result that all SLRs arrive at the Solar System conjointly. Top: The Solar System and NSM progenitor separated from each other by an area of high density ISM, e.g., the edges of a previous SN bubble or a superbubble, which is created when multiple CCSNe share the same shock front. Middle: The NSM progenitor explodes and pollutes the higher density area with its ejecta. Bottom: A CCSN explodes inside the bubble, pushing the NSM ejecta into the Solar System. CCSN and NSM ejecta arrive simultaneously at the Solar System location.}
    \label{fig:schematic}
\end{figure}

As seen in the previous sections, although each of the SLRs is produced in a different nucleosynthesis site, the density of all SLRs most likely increase synchronously in each given cell. This is consistent with the observation of the three SLRs detected in deep-sea sediments, \iso{53}Mn \citep[][]{Korschinek20}, \iso{60}Fe \citep[][]{Wallner16} and \iso{244}Pu \citep[][]{Wallner21}. Especially for the latter two, it has been shown in \cite{Wallner21} that they are both deposited in deep-sea sediments of the same depth, which leads to the conclusion that they arrived synchronously, even if they could have been produced in separate, independent nucleosynthesis sites. This could be interpreted in the following way, for the example of \iso{60}Fe from HMS and \iso{244}Pu from NSMs (Figure~\ref{fig:schematic}). An NSM explodes in an area of the Galaxy that is relatively close to the Solar System, however, it is separated from it by a region of high density, which could have been created by, e.g., a previous CCSN or a superbubble. Once the NSM explodes, the shock wave of the explosion, and hence the ejecta, will stop at the high density gas region, so none of the ejecta of this explosion event reaches the Solar System. If there is a HMS inside the region behind the shock wave (i.e., within the explosion bubble) of that NSM, that star will end its life in a CCSN. Once it explodes, it will create a second shock wave, carrying with it its CCSN ejecta. This second shock wave might just have enough energy to push the high density region (where the NSM ejecta were stopped earlier) towards the Solar System. Since the NSM ejecta are still conserved in this stopped high density area which is now pushed further by the subsequent CCSN, both ejecta from NSM and CCSN are conjointly pushed towards the Solar System. The result is a synchronous arrival of CCSN and NSM ejecta on Earth, hence \iso{60}Fe and \iso{244}Pu in Earth's inventory increase simultaneously, as observed.
In other words, NSM and other ejecta could \textit{``surf the wave''} of the CCSN explosion shock fronts. Since we have seen in Sections~\ref{sec:results:Pollution intervals}~and~\ref{sec:results:dominant prop mechanism} that all SLRs show a very similar $<\delta_\mathrm{SLR}>$, this propagation argument can also be made for the other rare nucleosynthesis site ejecta, \iso{53}Mn from SNIa, and also for the more locally deposited \iso{182}Hf from IMSs. In order to determine how often this ``surfing'' effect occurs, we calculate the fraction of CCSNe that sweep up significant amounts of SLRs in their blast wave: $97.35 \%$ / $99.94 \%$ / $99.89 \%$ / $99.81 \%$ of all CCSNe carry \iso{53}Mn,  \iso{60}Fe, \iso{182}Hf, and \iso{244}Pu, respectively, which means that almost all CCSNe contribute significantly to the propagation of all SLRs. To examine how this number is correlated with each nucleosynthesis site, we also consider a model where we lowered the frequency of NSMs by a factor of 100 (with $P_\text{NSM}=4\cdot 10^{-4}$, which would correspond to a theoretical gravitational wave rate of $\sim 18$ Gpc$^{-3}$ yr$^{-1}$, roughly four times below the current uncertainties of the LIGO/Virgo detections). In this model, the fraction of CCSNe that sweep up significant amounts of the NSM ejecta \iso{244}Pu decreases to $88.15 \%$, while the other values remain almost unchanged.
\subsection{Effects of varying yields and frequency of the events}
\label{sec:results:evolutionof SLRS:Yields} 
\begin{figure*}
\subfigure
{\includegraphics[width=\columnwidth]{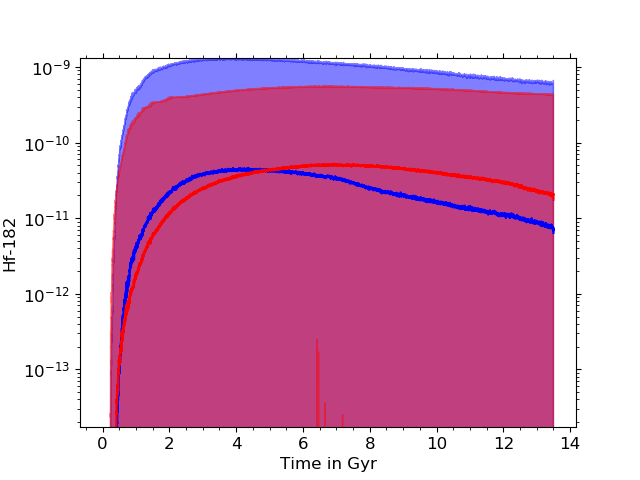}}
\subfigure{\includegraphics[width=0.49\linewidth]{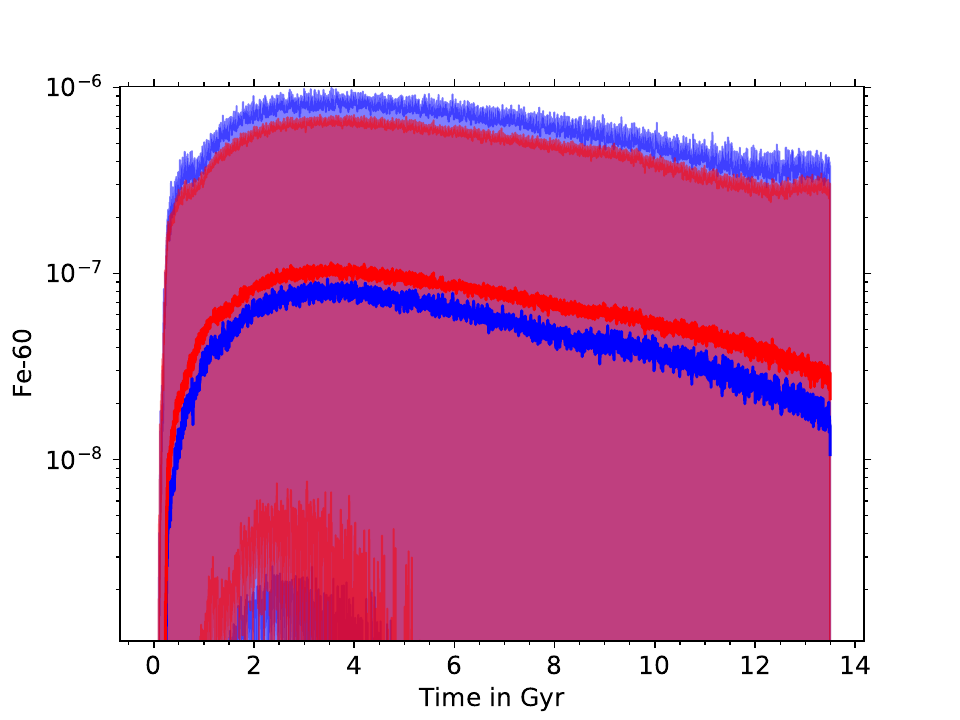}}
\caption{Effect of varied \iso{182}Hf (left) and \iso{60}Fe (right) yields. The solid lines represent the abundance medians and the shaded areas the full abundance distribution. The fiducial case is in blue and the case with the constant IMF-averaged yields is in magenta.}
\label{results:yields}
\label{results:evolutionof SLRS:HMSs:yields}
\end{figure*}

\begin{figure}
\includegraphics[width=\columnwidth]{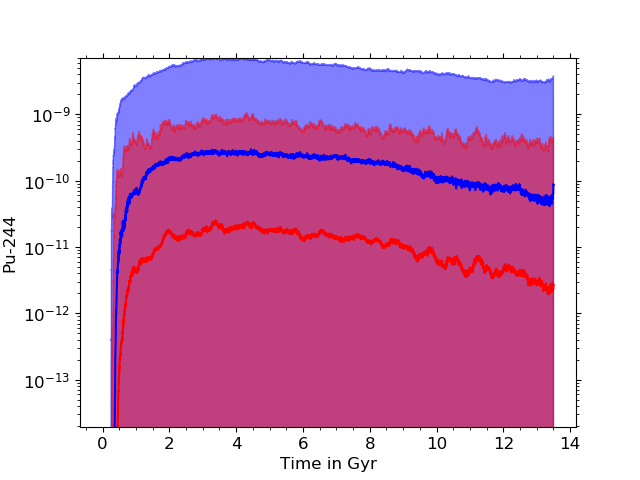}
\caption{Same as Figure~\ref{results:yields}, but showing the effect of decreasing the occurrence frequency of NSMs by a factor of $10$ (magenta), relative to the fiducial case in blue.}
\label{fig:evolutionof SLRS:NSMs}
\end{figure}

To test the impact of different yields and event frequencies on the results, first, we tested a case where the mass dependency of the yields was kept, but all their values were uniformly decreased by a factor of 10. The result confirms that a linear change of a factor of $\approx 10$ is reflected in the whole distribution. Second, we quantified the effect of the progenitor mass-dependency of the \iso{60}Fe and \iso{182}Hf yields on the ISM evolution of the SLRs. 
To this aim we changed the \iso{60}Fe and \iso{182}Hf yields from our fiducial yields from the literature (as described in Section~\ref{Sec:Model:Nucleosynthesis sites}), to a case where we took a constant yield for each progenitor equal to the IMF-weighted average over the entire ZAMS mass range.
\\
The results are shown in Figure~\ref{results:yields}. At the end of the simulation at time 13.475 Gyr, stellar births and deaths are almost in equilibrium, and a comparison made at this point allows us to minimize the impact of stochastic star formation or deaths. Using the constant yields, the \iso{60}Fe and \iso{182}Hf median increase by factor of $\approx 1.2$ and 3, respectively, at this late time of the evolution.
\\
The overall trend however differs: in the case of \iso{60}Fe the median calculated with the constant yields is always above the fiducial case, while for \iso{182}Hf the two lines cross at a time of roughly 4.5 Gyr.
The reason is that for the \iso{182}Hf ejected by IMSs, in the fiducial case the more massive IMSs eject more of this SLR, they also die earlier and therefore more \iso{182}Hf per IMS is ejected at earlier galactic stages, as compared to the model where all IMSs eject the averaged yield. At later stages, instead, the lower mass IMSs become more predominant, and in the case of constant yields, they eject more \iso{182}Hf than in the fiducial case resulting in a higher final \iso{182}Hf median. 
\\
The upper limit of the full abundance statistics also increases for \iso{182}Hf, however, the increase is lower, $\approx 1.2$, than that of the median value. This is due to a clumping effect more significantly present in the case of the fiducial yield. The more massive IMSs that contribute relatively more \iso{182}Hf are actually fewer than their lower-mass counterparts,  therefore their ejections are less homogeneously distributed than in the case where all IMSs all eject the same amount of \iso{182}Hf.
\\
For \iso{60}Fe from HMSs, the effect is weaker because the life time of the most massive HMSs compared to the least massive HMSs is not so different than the difference in life times of the most and least massive IMSs. In the fiducial yields case, the most massive HMSs eject far more \iso{60}Fe than the lower mass ones, which leads to a larger fluctuation of abundances (e.g., larger upper limit of the full abundance distribution) in the former case. However, the median of abundances is lower in the former case, because the majority of HMSs are (due to the IMF) at the lower mass end of the HMS mass spectrum. This leads to a lower abundance median in this case compared to the case where all HMSs eject the same, IMF-averaged yield of \iso{60}Fe.
\\
As discussed in \cite{Cote19b} and \cite{Yague21}, the occurrence frequency of stellar events influences how much time an SLR has to decay before a subsequent nucleosynthesis event. Since this frequency is governed by the rate of an event, this implies that also the event rates play a major role for the evolution of the SLR abundances. Although the rates of HMSs, IMSs, and SNIa have been determined by observations, the observational error bar on the occurrence rate of NSMs as suggested by gravitational wave detections is uncertain within a factor of $\approx$10. We illustrate the effect of changing the NSM rate on \iso{244}Pu.
\\
In Figure~\ref{fig:evolutionof SLRS:NSMs}, we present the evolution of \iso{244}Pu in our fiducial model versus a model with NSM occurrence rates reduced by a factor of $10$. The median of \iso{244}Pu ISM densities is reduced by a factor of $\approx 20$, i.e., it is two times lower than the case when the yields where reduced by a factor of 10. 
The reason is that a decrease in the rate results in less cells polluted by NSM ejecta per time step. A decreased yield, instead, results in the same amount of cells polluted per time step, but with a lower \iso{244}Pu density. Since the median is more sensitive to the amount of cells being polluted than the amount of \iso{244}Pu in the cells, the median value of the abundance is then lower in the former case.
\section{Conclusions} \label{sec:conclusions}
We simulated the evolution of four SLRs from four separate nucleosynthesis sites: \iso{53}Mn from SNIa, \iso{60}Fe from CCSNe, \iso{182}Hf from IMSs, and \iso{244}Pu from NSMs, over the lifetime of our Galaxy using a 3-dimensional GCE model. We conclude that: 
\begin{enumerate}
    \item The integrated (over the entire simulation volume) median ISM density of the four SLRs follows the Galactic star formation rate over time (with an individual delay for each SLR), but their density at a given location fluctuates strongly.
    \item \iso{60}Fe has the lowest delay since HMSs have the shortest lifetimes. Second is \iso{244}Pu since the lifetimes of two HMSs plus the coalescence time scale is still lower than the lifetime of IMSs (ejecting \iso{182}Hf, second longest delay time). \iso{53}Mn has the longest delay time since it always requires the longer-lived IMS in a IMS binary to die to produce that SLR.
    \item Even though the SLRs are produced at different sites, their arrival at a given location in the ISM often coincides, because their propagation is dominantly governed by the frequency of CCSNe.
    \item This explains why three different SLRs, \iso{53}Mn, \iso{60}Fe, and \iso{244}Pu, produced in three different nucleosynthesis sites (SNIa, HMSs, and NSMs) could be detected in layers of similar depth in deep-sea sediments, which suggest that they arrived conjointly on Earth. Our model can robustly reproduce these deep-sea detections.
    \item Neither assuming a PINBALL model, nor that all CCSNe explode as HNe strongly influences the overall evolution of SLRs.
    \item Reducing the occurrence frequency of NSMs reduces the median abundances of \iso{244}Pu. Reduced yields reduce the median abundances proportionally. Decreasing CCSN and IMS yields decreases the median abundances of \iso{60}Fe and \iso{182}Hf.
\end{enumerate}
Effects that still need to be addressed in future work are:
Those deriving from a different IMF, which would alter the lifetimes of stars and thus change the distribution of the pollution intervals, and from the existence of islands of explodability of massive stars \citep[e.g.,][]{Sukhbold16,Ertl16,Ebinger19,Curtis19}. The presence of failed CCSNe may reduce their dominance in being the major propagation mechanism for SLRs. Further, additional sources for the four SLRs should be addressed, e.g., rare sub-types of CCSNe as a source of \iso{244}Pu \citep[e.g.,][]{Fischer20}.
Also, further propagation mechanisms not currently included in our model could have an effect on the propagation of SLRs:
\begin{itemize}
    \item Diffusion/turbulence: \cite{Hotokezaka15} found that the discrepancies between early Solar System abundance and more recent deep-sea detection of \iso{244}Pu can be explained by a model that assumes a purely diffusion/turbulence-dominated propagation. This model was later expanded by \cite{Beniamini20}, who concluded that this approach could also applied to $r$-process elements. We did not include diffusion/turbulence in our model. If included, diffusion/turbulence might reduce (increase) the SLR densities in those cells with the highest (lowest) SLR abundances, which would lead to a reduced spectrum of SLR densities. However, these considerations would go beyond the scope of this work, where we wanted to highlight the ``surfing'' effect of SLRs. We plan to work on a comparison between CCSN-dominated, and diffusion-dominated propagation for SLRs in the future.
    \item Superbubbles: A large bubble in the ISM can be created by multiple CCSN explosions \citep[e.g.,][]{Krause18}. The effect of superbubbles is very difficult to estimate in our GCE model with limited spatial and time resolution. The implementation of superbubbles requires the implementation of thermodynamics/fluid dynamics \citep[as done in e.g.,][]{Vasileiadis13,Fujimoto18,Fujimoto20}, but this would go beyond the scope of this work. Nonetheless, more detailed effects that could be expected in hydrodynamical simulations can be estimated with our set of models with varying parameters, and all of our models show the surf effect.
    \item Galactic outflows: Galactic outflows could potentially built up a reservoir of gas outside the Galaxy that may be unaffected by GCE for a period of time \citep[e.g., as done with \textsc{Omega+} in ][]{omegaplus}, or may be enriched differently in hydrodynamical simulations \citep{vincenzo20}. The implications of this effect on SLR abundances are difficult to estimate in our model. Although SLR abundances (due to the absence of nucleosynthesis sites increasing the SLR abundances in the extragalactic reservoir) would potentially decrease exponentially in that extragalactic reservoir due to radioactive decay, gas being incorporated back from the reservoir into the Galaxy might slightly enhance the abundance of SLRs in the simulation volume. However, since our main aim for this work was to showcase the ``surfing'' effect for SLRs, we decided to use primordial inflow into the simulation volume as described in Section~\ref{Sec:Model:General setup}.
\end{itemize}

Future further detection of live radioisotopes in the deep-sea floor might provide further constraints on the propagation mechanism of SLRs \citep[e.g.,][]{Wang21a,Wang21b}.
This will also be of further interest for the GCE of $r$-process elements, since it is yet unclear whether the behaviour of different classes of elements (e.g., iron group and $r$-process elements) is coupled or not \citep[][]{Beniamini20}. A direct comparison between different or combined propagation mechanisms would therefore help to further confirm or rule out whether the propagation of SLRs and $r$-process elements is CCSN- or diffusion-dominated. 
Our study will also be extended towards a full cosmological zoom-in simulation of a galaxy \citep{Kobayashi11,vincenzo20}, which will allow for additional sub-galactic-scale mixing effects and mechanisms to be addressed (Wehmeyer et al., in prep.).
Additionally, our models can be applied to study the  abundances of SLRs at the time of the formation of the Solar System.
\section*{Acknowledgements}
The authors thank Anton Wallner for providing the deep-sea measurement data relevant for this work. We further thank Adrienne Ertel and Jesse Miller for constructive discussion with regard to the PINBALL model at the 2019 JINA-CEE Frontiers meeting.
This work is supported by the ERC Consolidator Grant (Hungary) funding scheme (Project RADIOSTAR, G.A. no. 724560). We also thank the COST actions ``ChETEC'' (G. A. no. 16117) and  ``ChETEC-INFRA'' (G. A. no. 101008324). MKP received funding from the European Union’s Horizon 2020 research and innovation programme under the Marie Sklodowska-Curie grant agreement No 753276. BC and BW acknowledge support from the National Science Foundation (NSF, USA) under Grant no. PHY-1430152 (JINA Center for the Evolution of the Elements) and Grant no. OISE-1927130 (IReNA).
CK acknowledges funding from the UK Science and Technology Facility Council (STFC) through grant ST/R000905/1, \& ST/V000632/1.
The work of AYL was supported by the US Department of Energy through the Los Alamos National Laboratory. Los Alamos National Laboratory is operated by Triad National Security, LLC, for the National Nuclear Security Administration of U.S.\ Department of Energy (Contract No.\ 89233218CNA000001). We thank an anonymous referee for valuable comments that helped improving this paper.
\bibliography{main}{}
\bibliographystyle{aasjournal}
\appendix
\section{Testing the resolution dependence}
\label{appendix:sec:resolution} 
The spatial resolution of our model is chosen to avoid a mass bias for newly born stars. In order to sample the entire IMF in the range 0.1 $\mathrm{M}_\odot \leq \mathrm{M} < $ 50 $\mathrm{M}_\odot$, a star-forming sub-cell has to contain at least 50 $\mathrm{M}_\odot$ of gas. This intrinsically limits the spatial resolution of the model since a sufficiently large number of sub-cells fulfilling this mass criterion has to be found at every time step. For this reason, we have chosen a spatial resolution of $(50 \mathrm{pc)^3}$. If the resolution was increased further, it would become more difficult to find a sufficiently large number of sub-cells to fulfil the mass criterion every time step. We can still test the spatial resolution dependence by decreasing it. 

\begin{figure*}
\includegraphics[width=\columnwidth]{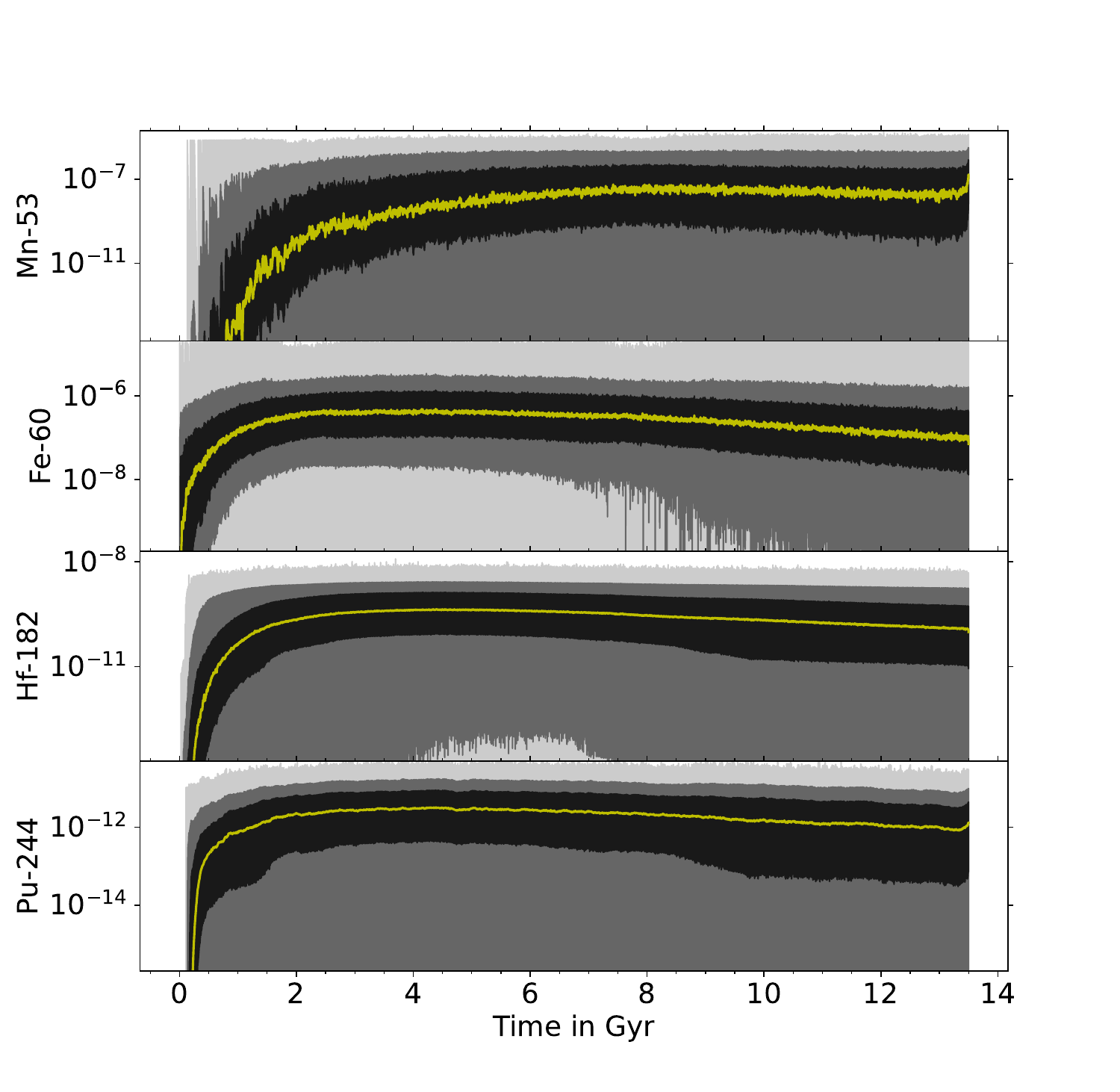}\hfill
\includegraphics[width=\columnwidth]{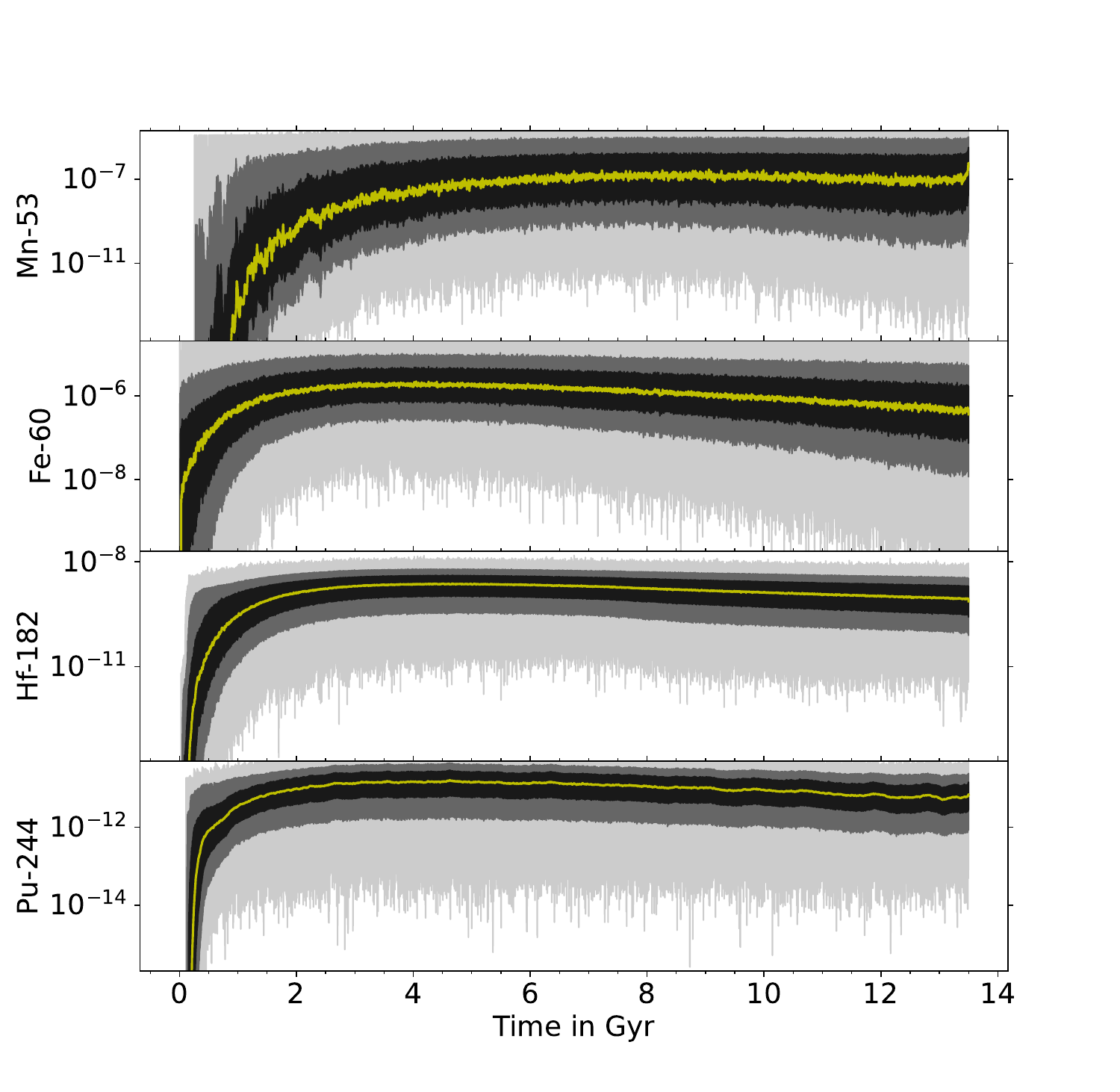}\hfill
\caption{Same as Figure 1, but with lower sub-cell resolution of an edge length of 80 pc (left panel), and 125 pc (right panel).} 
\label{fig:appendix:Spaceres}
\end{figure*}

To do this, we set up two models where we decreased the sub-cell resolution to $(80 \mathrm{pc)^3}$ and $(125\mathrm{pc)^3}$. The results in Figure~\ref{fig:appendix:Spaceres} show that the abundances have less variability. This is not surprising, as the lower the resolution, the more the results will converge toward those of a one-zone model, which would show a line for all abundances. Since the number of very low abundance sub-cells is strongly reduced in a model with lower resolution, the lower 100\% statistics boundary converges faster than the upper 100\% statistics boundary. This also slightly increases the median in the abundances, since it is strongly dependent on the number of sub-cells featuring a given abundance. This effect can be observed prominently in the \iso{244}Pu evolution in the right panel of Figure~\ref{fig:appendix:Spaceres}.

\begin{figure}
\includegraphics[width=\columnwidth]{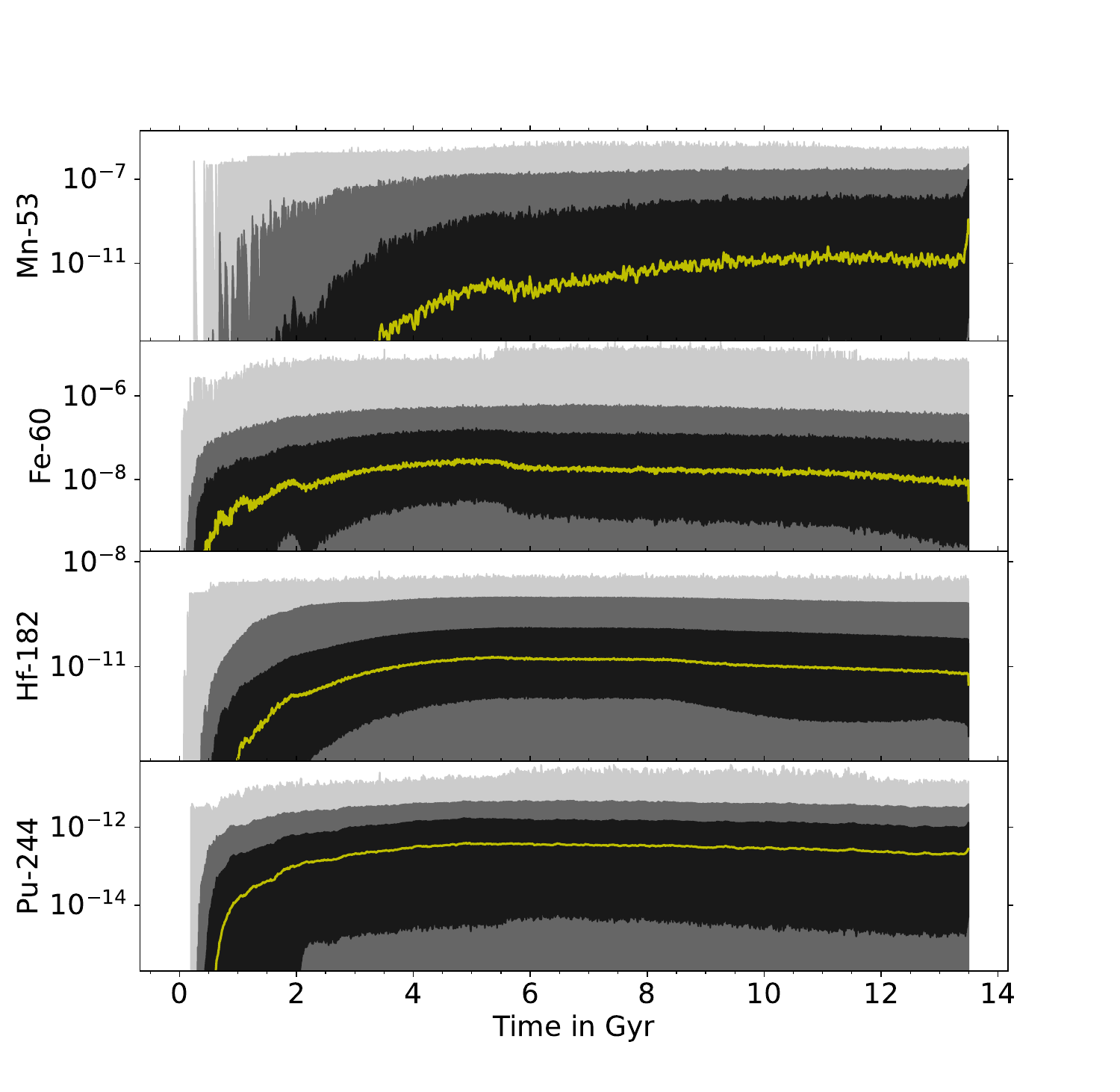}\hfill
\caption{Same as Figure 1, but with time resolution of 4 Myr per time step.} 
\label{fig:appendix:Timeres}
\end{figure}

For the time resolution, the time step of 1 Myr was chosen because the code cannot handle effects emerging from, e.g., hydrodynamics and cooling. After 1 Myr, the gas ejected from a SN explosion has completely halted and sufficiently cooled down, so it can be used to form the next generation of stars. Choosing time steps smaller than 1 Myr would require to model precisely the trajectory of a SN blast wave in the ISM within a hydrodynamic framework, which is not currently implemented in this model. As we have done above for the spatial resolution, we tested also the time-resolution dependence of the results by lowering it. We set up a model where we used 4 Myr time step. The results can be found in Figure~\ref{fig:appendix:Timeres}. In general, the upper 100\% statistics for all isotopes behave in a similar manner as in the standard case. However, the 95\% and 68\% statistics are significantly affected by the lowering of the time resolution. More nucleosynthesis sites explode per time step in the lower resolution model. This means that more cells are affected during each time step, which effectively leads to a stronger mixing of the entire volume, and less cells retaining lower abundance values. In other words, the likelihood of each cell to be affected by nucleosynthesis and mixing events at every time step is higher. This leads to the shrinking of the spectrum of the statistics, while the 100\% statistics seem to change only marginally. 

We caution that the tests presented here would need to be calibrated to some galactic properties (e.g., age-metallicity-relation, GCE of $\alpha$-elements), as done for the standard model presented in the main text, before they can be compared to the actual ISM.

\section{Increasing the amount of reflected material}
\label{appendix:sec:more geo} 
\begin{figure}
\includegraphics[width=\columnwidth]{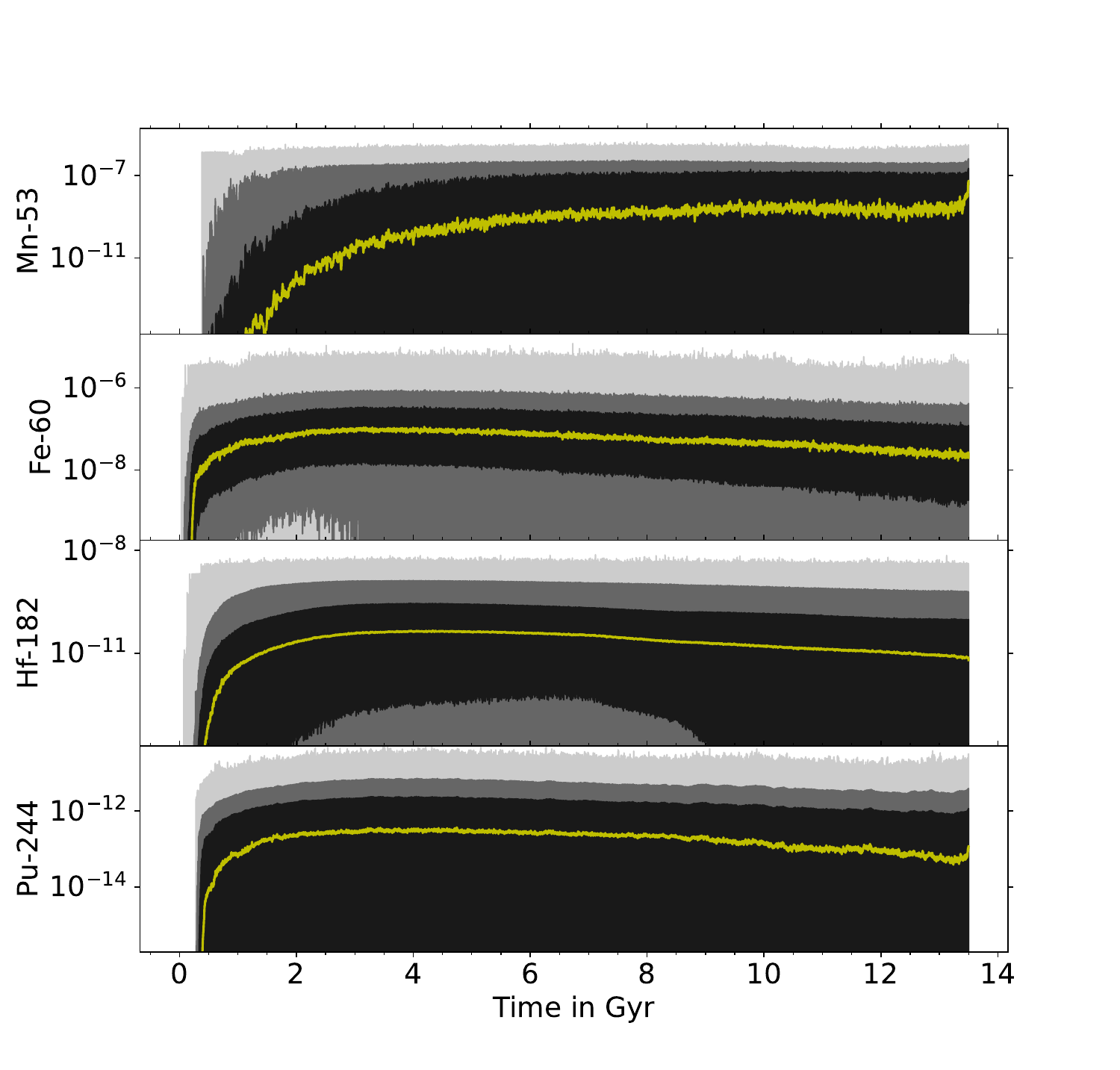}\hfill
\caption{Same as Figure 1, but with 50 \% deflected material instead of 1 \% as done in the PINBALL model.} 
\label{fig:appendix:GEO}
\end{figure}

The Sedov-Taylor approach assumes that all material  swept-up by a SN explosion is deposited solely on a bubble shell surrounding the SN explosion. The PINBALL model also explored in this work is motivated by the findings of \cite{Fry20}. When a magnetic field is present in the ISM around the SN blast wave, some material inside the outward SN shock front can be deflected in the backward direction. If that material reaches the opposite shock front, it might experience another change of direction caused by the magnetic field, causing the material to move through the inside of the SN shock front, like a pinball. \cite{Fry20} also conclude that the fraction of reflected material in a SN blast wave is dependent on the magnetic field strength and the size of the magnetic grains that condense behind the SN blast wave, and is difficult to constrain. To investigate the potential impact of such pinball remnant geometry,  we have introduced a PINBALL model in this work, by assuming that 1\% of the material swept-up by the SN blast wave remains inside the remnant bubble, and not on the shell. This number is motivated by \cite{Fry20} and we found that this choice does not strongly affect the evolution of SLRs. Here, we test whether a much larger fraction (50 \%) of deflected material would more strongly alter the abundance evolution. The results in Figure~\ref{fig:appendix:GEO} show that even such a high fraction of deflected material would not strongly affect the overall evolution of the SLRs. Therefore, this choice might affect SLR densities only locally in cells, but not their overall abundance statistics. This is in contrast to the HN model, which increases the overall size of the remnant bubble. However, we caution that such high fractions of deflected material would affect the ISM density distribution, which in turn would alter the star formation history due to the exponent of the Schmidt law (see Section~\ref{Sec:Model:General setup}) used in this model. This might yield results which are not congruent with Milky Way Galaxy properties (e.g., age-metallicity-relation, GCE of $\alpha$-elements and so forth).

\section{Robustness of the deep-sea detection fit}
\label{appendix:sec:robustness} 
To examine the robustness of our one sub-cell fitting approach applied in Section~\ref{sec:results:recentevolution}, we tested how much the mean squared distance of the logarithm of the abundances changes if the deep-sea detection data of the SLRs had a different time (x-axis) or abundance (y-axis) spacing. For the Standard model (with time and spatial resolution as in the main text), the value of this mean squared distance for the actual deep-sea measurements is $S = 4.54$, better than 83\% of fits\footnote{Determined from assuming the abundances follow a normal distribution with $\sigma$ equal to the uncertainty. Sampling the abundances to obtain the resulting distribution of $S$ allows us to compare a given $S$ with all possible fits. The goal is to provide a reference for the goodness of fit.}, with $S$ given by
\begin{equation}
    S = \frac{1}{N}\sum{\left(\frac{\ln R(t_i) - \ln Y_i}{\ln (Y_i + E_i) - \ln Y_i}\right)^2}.
\end{equation}
where $R(t_i)$ is the abundance of the run at time $t_i$, and $Y_i$ and $E_i$ the deep sea measurement and uncertainty, respectively, at time $t_i$.

\begin{figure*}
    \centering
\subfigure[Five times larger time spacing]{\includegraphics[width=\columnwidth]{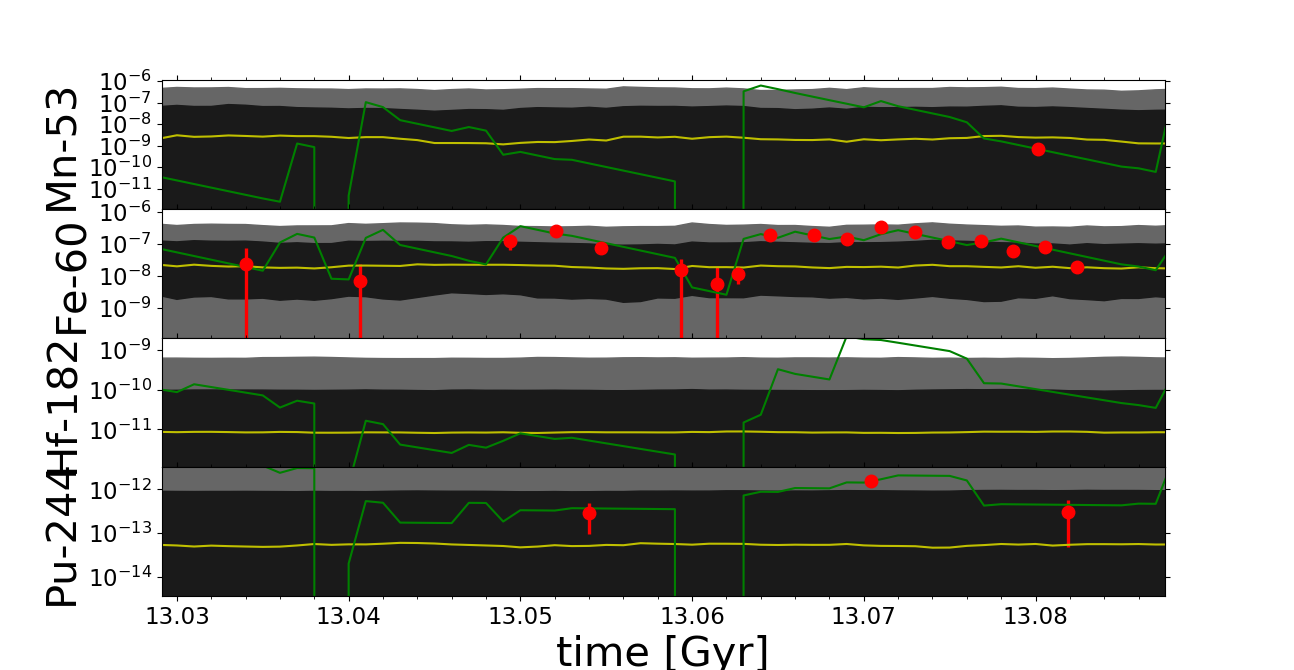}}\hfill
\subfigure[Two times smaller time spacing]{\includegraphics[width=\columnwidth]{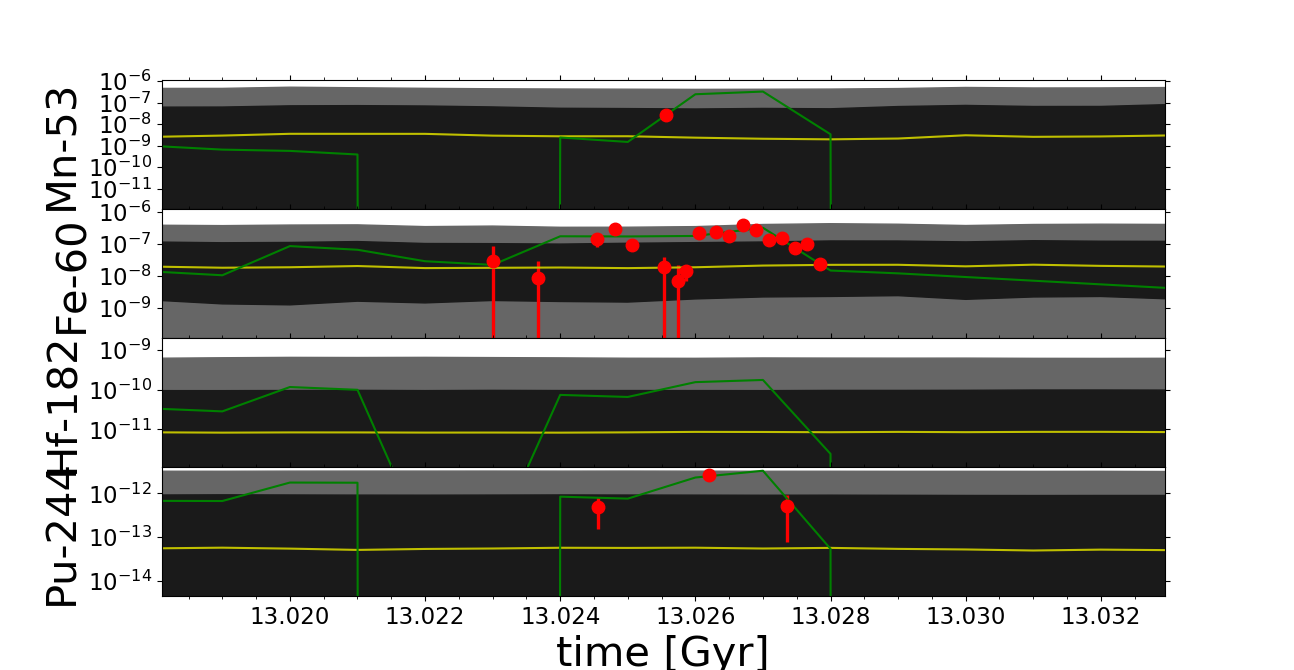}}
\subfigure[Ten times larger abundance spacing]{\includegraphics[width=\columnwidth]{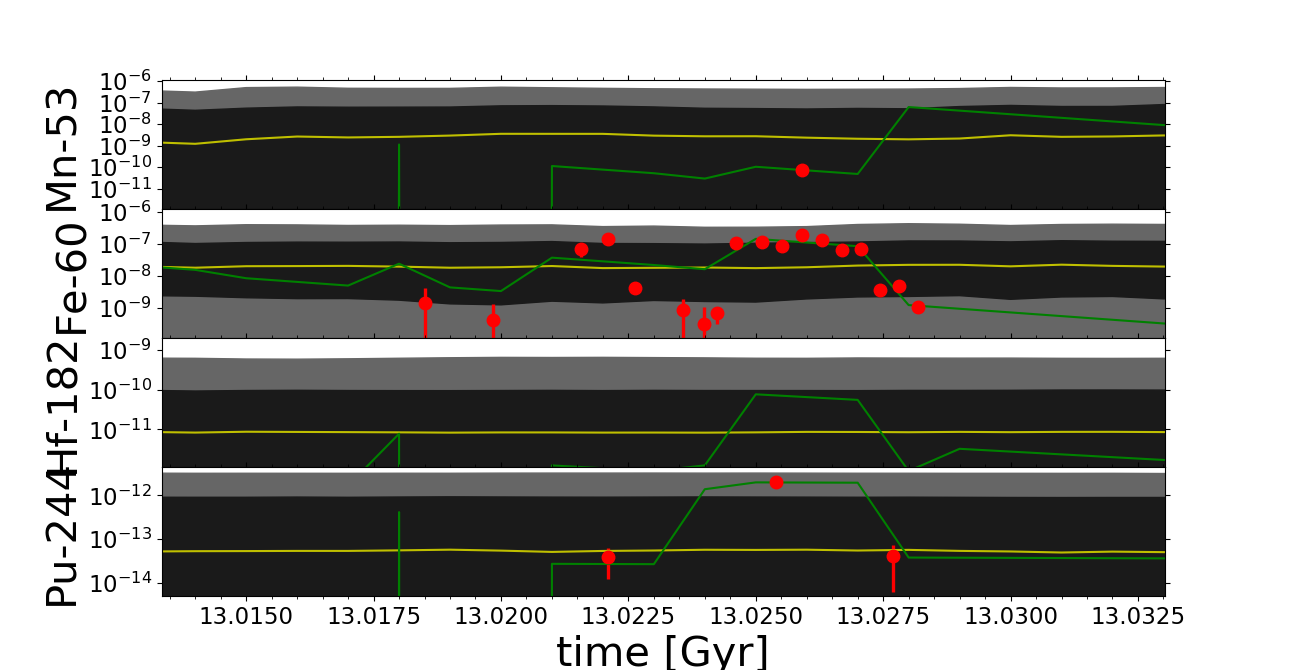}}\hfill
\subfigure[Ten times smaller abundance spacing]{\includegraphics[width=\columnwidth]{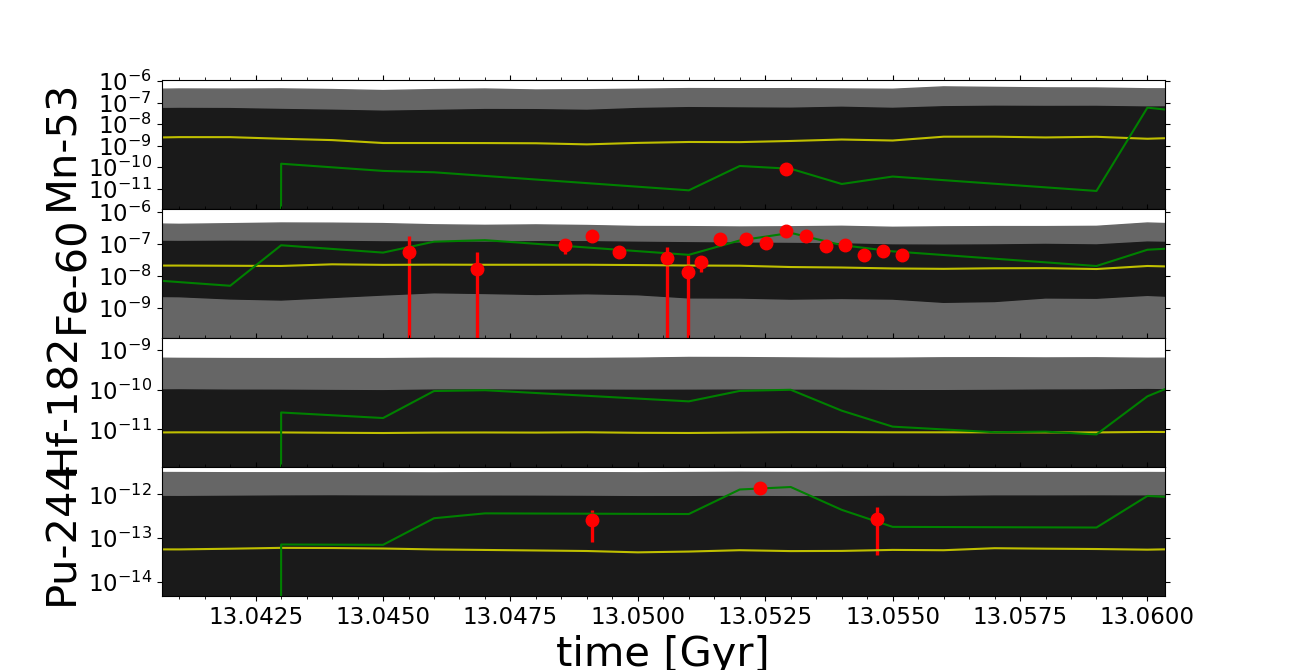}}
\caption{Same as Figure 2, with hypothetical different time and abundance spacing for the deep-sea detections.} 
\label{fig:appendix:spacing}
\end{figure*}

If the time spacing between the individual deep-sea detection data points was five times larger, it is easier to find a sub-cell that can match the detection data (see top left panel of Fig.~\ref{fig:appendix:spacing} for an example, with $S = 3.14$, i.e., better than 85\% of fits). This is because the sub-cells have much more options for the individual evolution between data points by either radioactive decay or the surf effect discussed in Section~\ref{sec:schematic_interpretation}. In other words, sub-cells have the opportunity to undergo the surf effect multiple times between two deep-sea detection data points when the spacing is larger, which increases the possible spectrum of SLR densities that sub-cells can exhibit. If the time spacing between detection data was smaller, the situation would be the opposite. Sub-cells have less possibilities for individual evolution (via radioactive decay or the surf effect), and thus the spectrum of possible abundances is lower. This makes it harder to find a sub-cell that fits the observations even if decreasing the time spacing between detection data only by a factor of two (see top right panel of Fig.~\ref{fig:appendix:spacing} for an example, with $S = 6.23$, better than 82\% of fits).
\newline
For the abundance spacing, we find a similar result. If the spacing between detected abundances was larger, we would have to find sub-cells that undergo stronger density fluctuations than those fitting the actual detection. This decreases the number of sub-cells that provide a fit (see lower left panel of Fig.~\ref{fig:appendix:spacing} for an example, with $S = 17.48$, better than 81\% of fits). Instead, the more we decrease the spacing between the detected abundances, the closer to a straight line the requirement for a single sub-cell to fit these detection is. Interestingly, especially after the right (latest) peak in the modified \iso{60}Fe data, these modified data points appear to scatter around the natural decay of that SLR. Since this area of the deep-sea detection data features many points, the weighting of the fit leads to preferring ``undisturbed'' sub-cells (i.e., sub-cells that only experience radioactive decay instead of the surfing effect), which makes it easier to find sub-cells to fit these hypothetical detection data (see bottom right panel of Fig.~\ref{fig:appendix:spacing} for an example, with $S = 2.89$, better than 86\% of fits).

\end{document}